# Urban Social Media Inequality: Definition, Measurements, and Application

**Agustin Indaco** (Economics, The Graduate Center, City University of New York)
**Lev Manovich** (Computer Science, The Graduate Center, City University of New York)

**Keywords:**

**Abstract:**

Social media content shared today in cities, such as Instagram images, their tags and descriptions, is the key form of contemporary city life. It tells people where activities and locations that interest them are and it allows them to share their urban experiences and self-representations. Therefore, any analysis of urban structures and cultures needs to consider social media activity. In our paper, we introduce the novel concept of social media inequality. This concept allows us to quantitatively compare patterns in social media activities between parts of a city, a number of cities, or any other spatial areas.

We define this concept using an analogy with the concept of economic inequality. Economic inequality indicates how some economic characteristics or material resources, such as income, wealth or consumption are distributed in a city, country or between countries. Accordingly, we can define social media inequality as the measure of the distribution of characteristics from social media content shared in a particular geographic area or between areas. An example of such characteristics is the number of photos shared by all users of a social network such as Instagram in a given city or city area, or the content of these photos.

We propose that the standard inequality measures used in other disciplines, such as the Gini coefficient, can also be used to characterize social media inequality. To test our ideas, we use a dataset of 7,442,454 public geo-coded Instagram images shared in Manhattan during five months (March-July) in 2014, and also selected data for 287 Census tracts in Manhattan. We compare patterns in Instagram sharing for locals and for visitors for all tracts, and also for hours in a 24-hour cycle. We also look at relations between social media inequality and socio-economic inequality using selected indicators for Census tracts.

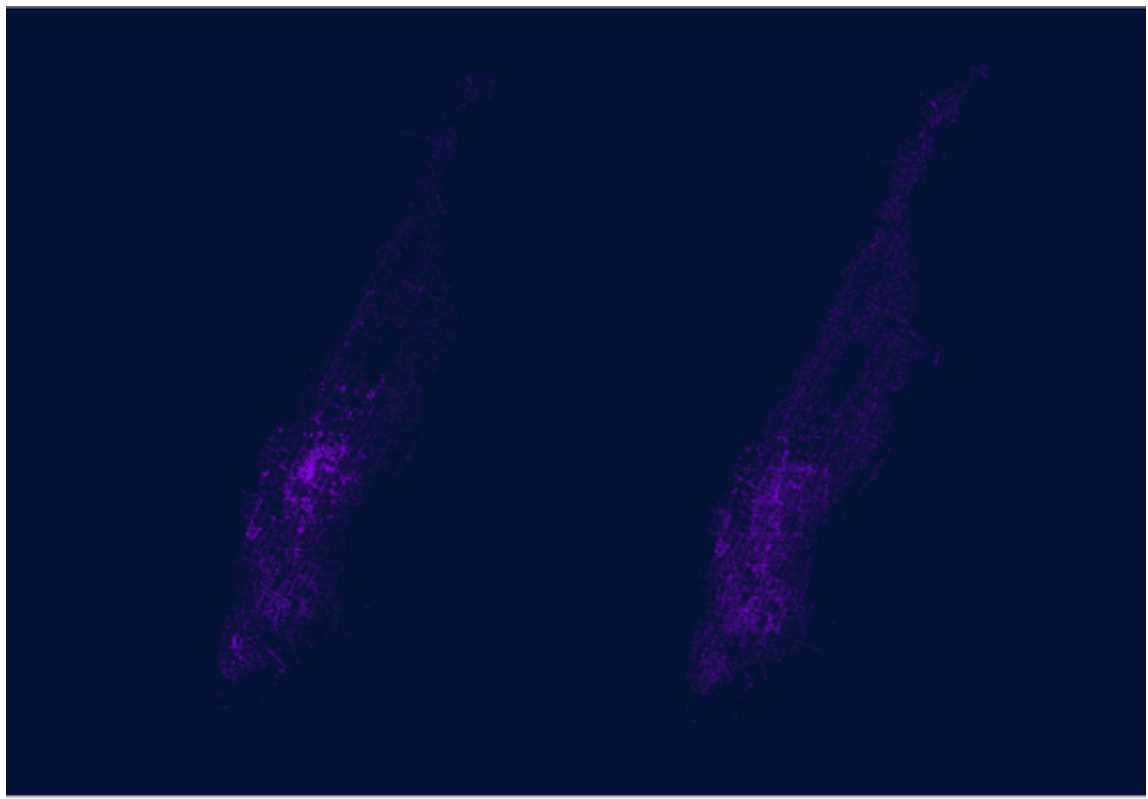

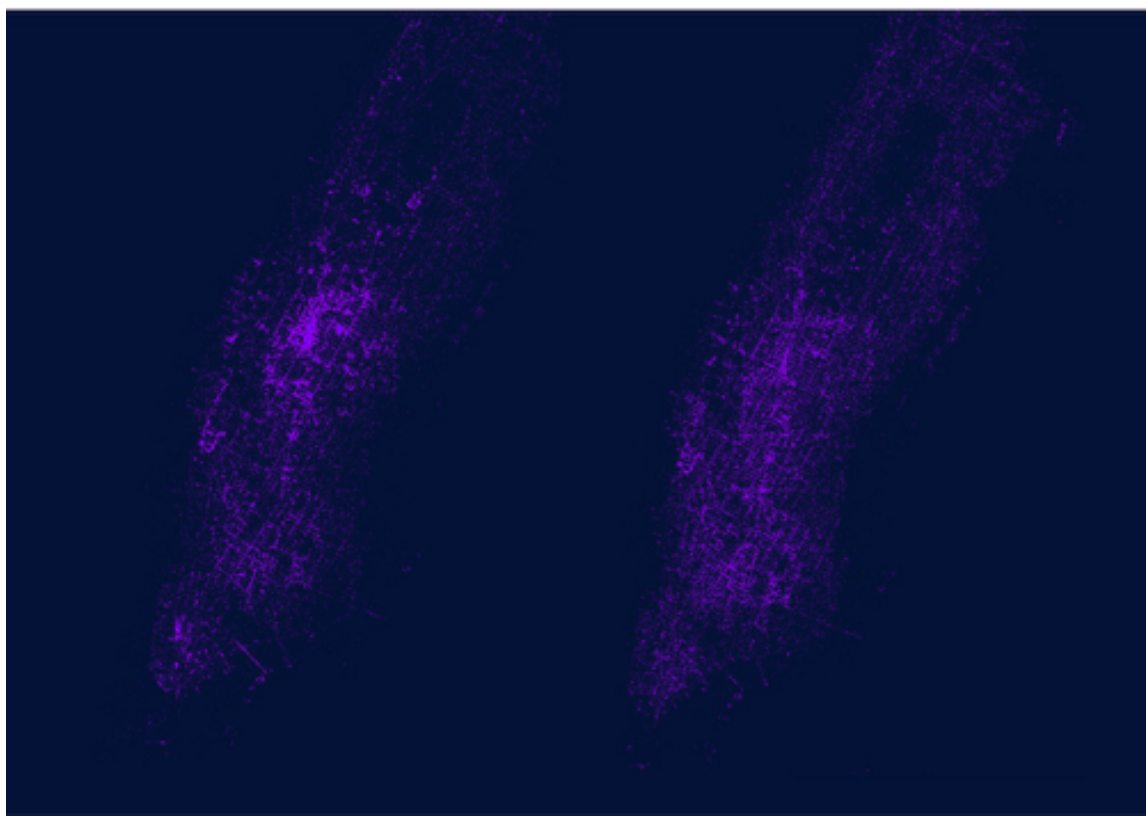

**Fig. 1 and fig. 2**.

# Introduction

Social media content shared today in cities, such as Instagram images, their tags and descriptions, is the key form of contemporary city life. It tells people where activities and locations that interest them are and it allows them to share their urban experiences and self-representations. Social media also has become one of the most important representations of city life to both its residents and the outside world. One can argue that any city today is as much the media content shared in that city on social networks as its infrastructure and economic activities.

For these reasons, any analysis of urban structures and cultures needs to consider social media activity and content. While the industry developed many concepts and measurement tools to analyze social media, these concepts and tools were not developed for comparative *urban* analysis. Therefore, we need to develop our own concepts that bridge the perspectives of urban studies and design and quantitative analysis of social networks that use computational methods and "big data."

In the last few years, one of the most frequently discussed public issues has been the rise in income inequality (Stiglitz, 2012; Piketty, 2014; Atkinson, 2015). But inequality does not only refer to distribution of income. It is a more general concept, and it has been used for decades in a number of academic disciplines besides economics, such as urban planning, sociology, education, engineering, and ecology. The quantitative measurement of inequality allows researchers to characterize a set of numbers or compare multiple sets, regardless of what the data represents. In addition to income inequality, we can measure inequality in wealth, education levels, social well-being, and numerous other social characteristics.

In our paper, we introduce the novel concept of *social media inequality*. We define this concept using an analogy with the concept of economic inequality. Economic inequality refers to how some economic characteristics or material resources, such as income, wealth or consumption are distributed in a city, country or between countries (Ray, 1998; Milanovic, 2007; OECD, 2011). Accordingly, we can define social media inequality as *the measure of the distribution of characteristics from social media content shared in a particular geographic area or between areas.*

An example of such characteristics is the number of photos shared by all users of a social network such as Instagram in a given city or city area. Another example is the number of hashtags – how many hashtags users added to the photos, and how many of these hashtags are unique. Other examples include average number of tweets shared by a user in a particular period; numbers of tweets shared per month, per week or per hour of a day; the proportions of tweets that were retweeted, and so on. Of course, we can compute and analyze features of content itself - for example, how many different subjects appear in the photos, and what are their proportions.  In fact, any metric of social media can be used to compare inequality in social media activity between areas - for example, number of likes, length of text messages, most frequent and least fre-quent words, number of unique topics, number of distinct photographic styles, image compositions, styles of video editing, and so on.

We propose that the standard inequality measures used in other disciplines, such as the Gini coefficient, can also be used to characterize social media inequality. We can also compare these measures between content shared on various social networks (Instagram, Twitter, etc.) in the same area or areas. We can do these comparisons for social networks where the main content is text (e.g., Twitter, VK), images (e.g., Instagram, Tumblr), video (e.g., YouTube), or combination of different media (e.g., Facebook, QZone, Sina Weibo, Line, etc.). Finally, we can also compare characteristics of shared content with various social and economic characteristics in the same areas, such as income, rent, the level of education, or ethnic mix.

The paper tests some of these ideas using a large dataset of Instagram images shared in Manhattan borough of New York City. This dataset, which we created for this study, contains 7,442,454 public geo-coded Instagram images shared in Manhattan during five months (March-July) in 2014. Among these images, 1,524,046 were shared by 505,345 city visitors; the remaining 5,918,408 images were shared by 366,539 city residents. Our analysis of the images shared by two types of users in this paper is inspired by the pioneering project *Locals and Tourists* created by Eric Fischer (Fischer, 2010.)

Comparing the location of images shared by visitors and locals (figure 1) gives us an intuition of the social media inequality concept. Figure 1 plots locations of random samples of images shared by tourists (left) and locals (right). Each sample contains 100,000 images. Figure 2 is a close-up of figure 1. We can immediately notice that in each case these, locations are not distributed evenly. Some parts of the city have many more images than other parts. These figures also suggest that the big proportion of images by city visitors are shared only in a few areas, while the locals share images in many more areas.

Note that we use the term "shared" rather than "captured" because Instagram allows users to share any image from their phone and not only those captured with the Instagram app. So users can upload images taken previously in other locations. However, since Instagram captures the geolocation and time when an image was shared (for users who allowed Instagram access to this data), the metadata of images in our dataset tells us about people's presence at particular place in the city at a particular time.

Visualizing the locations of shared images gives us an intuition for spatial social media inequality, but we need some measuring instruments to quantify such inequality. And what if we want to compare inequality not only in the number of images shared, but also in other characteristics we listed above (numbers shared per hour of the day, numbers of unique words in hashtags, etc.)? As these characteristics multiply, the need for quantitative measurements becomes stronger. Our paper proposes such measurement instruments and tests them using a few characteristics from images and accompanying metadata in our dataset.

In principle, we could also study social media inequality between individuals living in a city. This will be similar to how economists measure income inequality by comparing people's in-come, rather than the average income by area. However, to do this would require disclosing the identity of the individual behind a social media account, and thus going against privacy norms accepted in most countries today. At least until now, social networks such as Instagram, Twitter, Facebook and others have allowed researchers to download content shared by their users, but they do not disclose any user information beyond what users made visible on their account pages.

While the U.S. Census collects data on individuals, it only reports the data aggregated by geographic areas at different scales. We follow a similar logic in our analysis of spatial social media inequality by dividing a city into hundreds of small areas and aggregating characteristics of social media content shared in each area - as opposed to comparing individuals to each other. The way we measure social media inequality is comparable to how Branko Milanovic defines one of the measures of global economic inequality (Milanovic, 2006, Concept 1). This measure uses countries as the units of observation. Milanovic does not directly compare the income of people worldwide. Instead he compares average income across different countries to calculate global inequality. In our case, the Census tracts are our units of observation. We aggregate social media characteristics at the tract level in order to analyze social media inequality across all of Manhattan.

Social media content shared in a given area may combine contributions from different kinds of users: people who reside in this area, people who live in different parts of the city or in suburbs but spend significant time in this area for work during weekdays; international or domestic tourists visiting a city; companies located in this area, and so on. Together, the content shared by all these users create a collective "voice" of a particular area of a city. A city as a whole can be compared to an orchestra of all these voices (although, of course, they are not necessary performing the same composition.) Applying the concept of inequality to a collection of these urban voices can give us new ways of understanding a city, and provide an additional metric for comparing numerous cities around the world.

We can separate various types of users in each area, comparing social media use between locals, commuters, occasional visitors from other parts of the city, one-time tourists, and companies. The social media contributions of all agents of a particular type forms its own "social media layer" over a city, and comparing these layers inside one city, or between cities, can be very revealing. We can also compare and quantify the differences in the distribution of trajectories of people as reflected in the locations of images they take during particular intervals. This idea is inspired by Fischer's interactive visualization of movements of hundreds of millions people worldwide who shared their geo-tagged Flickr photos between 2005 and 2015 (Fischer, 2015).

Social media inequality as we define it refers to the unequal distribution of social media content and its metadata and their characteristics in any type of geographic area – a city, a region, a country, or any other type of area. However, as Fischer's maps show visually, the density of social media contributions in larger cities is much higher than in non-urban areas, which makes these cities particularly convenient areas of study. We think that our proposed measurements of social media inequality can be useful for urbanism studies, urban planning, urban design, public administration, economics, and other professional and academic fields. While researchers in the fields of social computing, spatial analytics, and "science of cities" have published many quantitative studies analyzing urban data of many kinds (Batty, 2013; Goldsmith and Crawford, 2014; Townsend, 2014; Pucci et al., 2015; Ratti et al., 2006), a significant portion of this analysis cannot be approached without having a degree in computer science. In contrast, social media inequality measurement is a concept that is easy to understand and also easy to compute.

The locations of social media contributions reflect the presence of people in a particular part of a city at a particular time. However, in comparison to pure location data captured by mobile phones or other body sensors, social media images are much more than simple coordinates and time stamps. The content of these contributions can also tell us what people find interesting and how they are spending their time. Therefore, mapping and measuring inequality in the distribution of social media posts can help us understand how social, economic, and urban design characteristics of cities influence life patterns and the overall "dynamism" and "vitality" of a city.

For example, a city tourist office may be interested in knowing which places are popular among visitors, which places they avoid, and how such characteristics of their city can be compared to similar characteristics of other cities with similar number of tourists. Quantifying the characteristics of spatial distribution of visitor activity using the proposed inequality measures would allow for such comparisons. In contrast, "top lists" (i.e., most frequently visited places) or other tourist statistics, such as average number of days that tourists spend in a city, miss most of the relevant information – such as the activity of visitors in all other areas that do not correspond to distinct landmarks.

Public policy officials could use social media inequality measurements in deciding what sort of laws and regulations to implement in areas such as transportation and zoning laws. By calculating social media inequality before and after some policy was implemented, we have additional measures to understand if this policy had an effect and to quantify the size of this effect.

Researchers have never observed perfect equality in any natural, biological or social system or population. In using the term "social media inequality," we are not suggesting that the goal of urban planners or city administration should be to reduce differences in social media use between various areas to a minimum, or to some optimal level. If people are sharing the same amount of social media in every area of the city, it means that this city does not have any centers or attractions that stand out, or places where many people gather. In terms of modern housing, large American-type suburbs with the same density of houses and same demographics of families and income would probably generate least amount of social media inequality. Today such suburbs are common around the world, from Mexico to China. Given the wide criticism of this classical suburb type, we can assume that *some level of spatial social media inequality is desirable*. In this case, inequality stands for variety and differentiation while complete equality stands for sameness and lack of variety.

But is extreme social media inequality a good thing? For example, do we really want all people living in a city to spend their weekends in a single place? There are certain situations where reducing extreme spatial social media inequality would be desirable. For example, if city authorities find that most tourists' social media activity is concentrated in just a few areas surrounding only a few landmarks (like Times Square in New York City), they can change the way the city is promoted to visitors to diversify where tourists go, what they look at, and what they experience. Being able to quantify inequality of social media would allow for better planning and evaluation of such changes.

Formulated as a type of spatial analysis, our study compares the parts of the city that attract more people and generate more content shared on social media networks and thus are "social media rich" with parts of the city that are "social media poor." What is the relationship between such social media rich and social media poor areas? Is social media inequality larger or smaller than economic or social inequality in the same areas? Does social media inequality increase worldwide, similar to how economic inequality has been growing recently? Which parts of the world have the highest social media inequality and which are the most equal? Although our analysis is focusing on one part of a single megacity (i.e., Manhattan in New York City), it can be expanded to consider hundreds of cities around the world to consider such questions.

In summary, we see the following as the main contributions of our paper:

1) We introduce a novel concept of "social media inequality." We define it as the unequal distribution of quantitative characteristics of user-generated content shared by people in a geographic area, a number of areas or a set of individuals. The examples of such characteristics are the number of shared images or messages, the numbers of hashtags and words used in posts or image descriptions, distributions of subjects and photographic techniques used in photos. The concept of "social media inequality" is different from the concept of "digital divide" because it refers to content people shared on social networks, rather than the access to these networks or Internet as a whole.

2) We show that popular quantitative measures of inequality such as Gini coefficient can be used to measure social media inequality.

3) We propose to use two types of social media characteristics in measuring social media inequality. The first are the descriptive statistics *about* shared content and the users who share it. The examples are the number of posts in a particular area, the number of unique users posting in an area, numbers of posts per user, per hour, per day of week, etc. In addition to such numbers, we can calculate and use statistics such as mean, variance, and so on. The second are characteristics of social media content *itself* that can be measured by computers. Researchers in the fields of social computing, computer vision, computer multimedia, and text analytics use hundreds of such characteristics (referred to as "features") of text, images, video and other types of media, and we can also use them in measuring social media inequality. The examples of such characteristics for text posts or descriptions of shared images and video are numbers of characters and proportion of nouns and verbs. For photos, it is now possible to determine with good degree of accuracy the objects and type of scene in a photo or video, and the kinds of photographic or video techniques used (Schifanella, Redi, and Aiello, 2015). Accordingly, in our study we look at examples of both types of characteristics (volumes of shared images, and number of total and unique hashtags).

4) We can analyze relations between social media inequality and economic and social inequality of populations in the same geographic areas. To illustrate this, we compare characteristics of Instagram images in Manhattan with selected data from the U.S. Census for 287 separate areas of the city (i.e., Census tracts). To the best of our knowledge, this is the first time social media activity is analyzed together with socio-economic characteristics of areas on such a hyperlocal scale.

5) We can expect that sharing patterns and the resulting social media inequalities will differ depending on the type of user. To illustrate this, we compare the spatial distributions of Instagram images and associated tags shared by visitors and locals in Manhattan. Given the economic and social importance of tourism today for the well-being and future of many cities, being able to quantify visitors' social media activity in detail is a particularly important application of the social media inequality concept.

6) In addition to the spatial inequality, we can also analyze temporal inequality of social media use. Accordingly, we consider patterns of inequality in relation to time of day, and days of the week.

Our paper is organized in five parts. The first part presents related research. In the second part, we discuss popular quantitative measures of inequality and provide examples of how the concept of inequality is used in various academic fields. In the third part, we analyze social media inequality for different parts of Manhattan. This case study uses a dataset containing 7,442,454 geo-tagged images shared on Instagram in Manhattan over 5 months in 2014. We compare spatial distributions of image locations for visitors and locals using a number of alternative inequality measures. We also look at temporal inequality - specifically, differences between numbers of images shared between months, days of a week, and hours in a 24-hour cycle. In the fourth part, we analyze relations between social media inequality and socio-economic inequality using a number of indicators from the U.S. Census for all 287 tracts in Manhattan. In the fifth part, we demonstrate how other characteristics beside volume can be used to analyze social media inequality using as examples two characteristics: number of hashtags assigned by users to Instagram images, and unique number of hashtags.

# 1. Related research

In order to map and quantify social media inequality in a city like New York, we can divide the city into many small areas and then compare measurements from these areas. We can for example use a rectangular grid to define such areas. However, because we want to compare social media inequality with social and economic inequality using the U.S. Census indicators, we adapt the same spatial divisions used by the Census in collecting and reporting data. For our study, we chose the type of area that the Census calls "tracts." There are 287 census tracts in Manhattan.

Each area has on average 3,000-4,000 people living there and it occupies 0.36 square km on average.

We aggregate selected social media characteristics of images shared inside these tracts and their metadata. Since the U.S. Census reports various socio-economic indicators for the same tracts, we compare social media characteristics and selected socio-economic indicators for the corresponding tracts. (In this part of analysis, we leave out contributions by visitors and only consider 5.9 million images shared in Manhattan by remaining users.)

To the best of our knowledge, the analysis offered in this paper is the first of its kind. We are not aware of any other paper that studies the relation between social media sharing and economic and demographic characteristics of geographic areas – i.e. looking at social media and socio-economic inequality together.

It is important to distinguish our work from both the extensive studies of "digital divide," and also statistics about social media use frequently reported by various companies and research centers, such as Pew Research Center (Duggan et al., 2015). The concept of digital divide is attributed to Allen S. Hammond IV and Larry Irving who analyzed the inequalities of online access across the U.S. in the middle of the 1990s (Hammond, 1997). Since then, there have been many studies on the disparity of online access, and its connection to social and economic inequalities (for example, van Deursen and van Dijk, 2011). Researchers have also analyzed differences in access both within particular countries and between countries.

Pew Research Center conducts surveys about use of internet, mobile phones, and popular social media services (Facebook, Twitter, Instagram, Pinterest) by people from different demographic groups divided by age, gender, income, education, ethnicity, and other characteristics. However, most of these surveys do not differentiate between different kinds of social media use: browsing, commenting, sharing other people's content, sharing new original content, etc. Additionally, since Pew Center surveys use small population samples (in the order of two thousands from across all U.S. states), they do not contain information on regional differences between areas.

Both digital divide studies and surveys of social media use implicitly follow the classical concept of economic inequality: comparing differences in access to some resource. In our case, rather than looking at patterns of *access* to a particular digital resource (i.e., Internet as a whole, blogosphere, one or more social networks or media sharing services, or mobile apps), we look at how people *contribute* to this resource. In the case of Instagram, for example, these contributions include images, their descriptions, hashtags and location names. While people certainly also frequently share existing online images created by others, our analysis of over 15 millions of Instagram photos shared between 2012 and 2015 by people in 16 global cities including NYC suggests that most are photos or screenshots people captured themselves with their mobile phones (Manovich, 2016).

However, to interpret the spatial patterns of social media sharing in a city among locals, we certainly need to consider a contemporary version of possible digital divide: owning a mobile phone with a good camera and the appropriate data subscription plan that allows sharing many photos on social media services. A recent Pew Internet survey reported that across the U.S., 21% of adults use Instagram. The percentage among Black and Hispanic online adults was higher than among White online adults; there were also more female than male users. The report did not find any big differences by groups in various income levels (Duggan et al., 2015). While this survey suggests that as a whole, the digital divide of social media and mobile phone use in the U.S. has practically disappeared, these patterns may be different in particular urban areas. For example, while our study does not look at individual users, when we aggregated numbers of shared Instagram images for every tract, we found a very large difference between areas of Manhattan with different income and ethnicity.

The presence of location information in shared social media allows us to analyze spatial inequalities at levels of granularity not captured by small scale surveys. We can also take advantage of another information automatically captured when a user shares a photo: the date and time, recorded as hour, minute and second. Accordingly, in this paper we analyze inequalities across dimensions of space and time. Both dimensions are equally important. Census surveys only capture the information for people who live in each area. But social media content is also shared by people who spend working hours in this area, come for entertainment and leisure activities, or visit the city for only a few days or hours. And in a borough such as Manhattan, the number of people who commute there for work every weekday is almost as large as the number of residents. Specifically, 1,636,268 people reside in Manhattan according to the 2014 Census estimate (U.S. Census Bureau, 2014). But during weekdays the population increases to 3.1 million (Roberts, 2013). And we also should not forget the tourists: in 2013, 54.3 million domestic and international tourists visited New York City (NYC statistics, 2015).

There is already a large body of quantitative research that uses content shared by people and their other digital traces to analyze patterns in urban social media, mobility, tourist activities, and also large scale movements of people across countries and around the world. The emergence of social media along with the ability to freely download large data samples from many popular networks has not only provided researchers across all fields with new information to help tackle established questions, but also led to the emergence of new research topics. For example, Sobolevsky et al. (2015) use three different datasets to quantify the ability of different cities in Spain to attract foreign visitors: bank card transactions, geo-tagged photographs and tweets. Salesses et al. (2013) use Google Street View images and crowd sourcing to measure people perceptions of safety, class, and uniqueness in different cities, and then relate these measures to homicide rates in these cities. They show how such measures of "perceptual inequality" capture information not contained in economic measures. Shelton et al. (2015) use geo-tagged social media data to study intra-neighborhood segregation, mobility and inequality. Using locations of tweets, they analyze city-wide mobility patterns of people from different socio-economic groups and show how popular spatial imaginary of a city divide between classes is not accurate.

In another example, Yang et al. (2014) use data from a location-based social networking service in order to study mobility patterns of natives and non-natives in Chinese cities. The authors find that non-natives have more sparse and heterogeneous mobility patterns, whereas natives tend to remain close to their home and workplaces. This is the only paper we are aware of that uses the same tools as our paper to characterize social media patterns: Gini coefficient and Lorenz curve. However, these tools are used by authors in a limited way and only quantify urban mobility heterogeneity in the cities they analyze. Our paper posits a general concept of social media inequality that can be applied across different spatial scales, with a variety of social media characteristics, and variety of ways to measure it besides Gini coefficient. Furthermore, the authors did not study social media inequality across different dimensions (i.e., space, volume, content, and time). We think that using such multiple dimensions to conceptualize a city, quantify inequality patterns, and compare them is the key advantage of a general social media inequality concept as defined in this paper.

The term "social media inequality" itself was used until now only in a single paper (Schradie, 2012) to the best of our knowledge. But the use of this term by Schradie is very different from what we propose in this paper. Schradie studies whether there is a socio-economic production gap in blogging, i.e.: if higher-income individuals are more prone to blogging relative to those with lower incomes. The author argues that blogging requires many resources, and this can produce a divide driven by socio-economic inequality. The paper finds that this is in fact the case: socioeconomic class inequality is the most persistent characteristic that helps explain differences in blogging, more so than race and ethnicity. The concept of social media inequality in this paper is centered on the idea that socio-economic differences help explain active use of social media.

This work is very relevant to our own analysis. However, the meaning of "social media inequality" in our paper is quite different. We use it to refer to the distribution of social media sharing across a spatial area or areas. The analysis of the possible relations between such distributions and distributions of income, wealth, education, and all other socio-economic variables can be certainly also undertaken, as we demonstrate below. But as defined and detailed in our paper, "social media inequality" is first of all a concept and a set of measurements tools for spatial theory, urban studies, and urban design. We believe that patterns in contemporary culture including production of content by individuals are driven by lots of factors, and not only by socio-economic variables. We also believe that these patterns vary significantly from place to place around the world, and that we should first map them in sufficient detail to understand their differences. The concept of social media inequality is one such instruments designed to compare patterns in social media in as many locations as we want.

# 2. Measuring Inequality

## 2.1. Use of inequality concepts in different disciplines

The most common quantitative measure of inequality used today – the Gini index – was introduced over hundred years ago by Italian statistician and sociologist Corrado Gini (Gini, 1912).

In economics and sociology, inequality measurements quantify how particular characteristics are distributed among members of a group, or between many groups. The examples of such characteristics include income, wealth, health, education, etc. If we think of these characteristics as resources, inequality measures quantify how even or uneven a certain resource is distributed among a group. For example, if everyone in a group has the same wealth, we have complete equality (Gini index = 0). If, on the other hand, one person has all the wealth and everybody else has nothing, we have highest possible inequality (Gini index = 1).

Many other research fields also developed concepts and measurement techniques to analyze inequality in distributions of artifacts and/or their properties specific to their fields. In economics, the concept has been employed primarily to study income, wealth or consumption inequality (Heshmati, 2004). For a long time, economists have been interested in understanding how income and wealth of an economy is distributed among the population of a certain city, country or the entire world (Milanovic, 2012a). Inequality is currently one of the most debated subjects in the economic literature, with many discussions about the growing income inequality in developed countries, its effects on economic growth and what it means for our societies. In 2011, two economists published a paper that showed that greater income equality increased the duration of countries' economic growth periods more than many other factors such as foreign investment, low foreign debt, free trade or low government corruption (Berg and Ostry, 2011). In his 2012 State of the Union speech, President Obama called inequality "the defining issue of our time" (State of the Union Address, 2012).

But the concept of inequality - or related concepts that characterize data distributions - are also commonly used in many other disciplines. In the fields of urban design and urban planning, researchers study population density and other characteristics of geographic areas, measuring their spatial distribution and spatial structure. In order to study such patterns, geographers and urbanists have developed a wide array of methods. One approach, called point pattern analysis, compares distances between observed units of development with theoretical distributions. For example, Thomas (1981) compares the average nearest-neighbor distance to that of a purely random location to identify possible centers or subcenters within a city.

A more recent approach is based on fractal dimensions. Fractals are patterns of elements that are "self-similar" at different scales (Mandelbrot, 1982). Given that cities are amorphous, irregular and do not have a clear inner organization (Caglioni and Rabino, 2004), fractal geometry provides a useful approach in studying spatial structures. In fractal analysis, the first step is to generate a large fractal and then chose a geometric operation that transforms the initial fractal into smaller self-similar patterns of itself. This process is done iteratively, exponentially increasing the number of figures and eventually representing the overall spatial pattern of a city. Once the fractal simulation represents the city being studied, urbanists can calculate several statistics based on it; for example, counting the number of occupied points that lie at a certain distance from other occupied points to study the spatial distribution.

But the most widely used method to study spatial structure among urbanists is spatial entropy (Bhatta et al., 2010). This method was introduced in the 1970s (Wilson, 1970; Batty, 1974; Batty, 1976). The main advantage of this method is that it is not affected by the number of sub-areas in the region of study (Thomas, 1981; Tsai, 2005) and can be applied to analyze the distribution of any particular characteristic in a spatial unit or metropolitan area. Relative entropy values range between zero and one; values close to zero indicate that the distribution is highly concentrated, and values close to one indicate that the distribution is highly dispersed. Spatial entropy is widely used in studying urban sprawl (for example Yeh and Li, 2001). However, one disadvantage of entropy measures is that different urban structures (for example, monocentric, polycentric and decentralized) will have the same values of entropy as long as their population distribution are the same (Tsai, 2005). Since entropy measures do not capture such differences, this limits their application in spatial analysis.

The study of inequality is also a key topic in sociology, where one of the goals is to understand the relationships between peoples' social situations and their professional opportunities. In many societies, distribution of resources is organized according to meritocratic principles. But how does this work in practice? Sociologists study if the distribution of resources is biased towards particular social groups. The most common types of inequality studied are gender, racial, ethnic, age, and health inequality. For example, groups that have different gender, ethnicity or age maybe subjected to discriminatory hiring and pay practices. As an example, the 2014 Global Gender Gap Report by the World Economic Forum scored 144 countries according to their level of gender inequality. The most equal countries are Iceland and Finland, with a female-to-male ratio above 0.84); the most unequal are Yemen and Pakistan, with ratios below 0.05. USA was in the 20[th] place from the top (ratio = 0.74) (WEO, 2014).

Ecology is also using inequality measures to compare number of species in different habitats. But instead of the term "inequality," ecologists more commonly use the term "diversity," or its equivalents "species evenness" or "species richness" (Wisley and Stirling, 2006; Whittaker, 1972). Ecologists developed a number of indexes to measure diversity such as counting the number of different species present in one area, or counting how many individuals of each species is present in the area. Another type of diversity indexes is called "functional attribute diversity" and takes into account the traits of each species present (Walker et al., 1999). The concepts and indexes developed by ecologists to quantify the differences between habitats gives us many very useful tools for comparing social media use between city areas. In particular, we think that the idea of functional attribute diversity is quite relevant if we intend to compare multiple characteristics of social media content shared in a number of areas.

In linguistics, the concept of inequality has also proven to be useful in understanding the functioning and evolution of language, the use of particular words, and other questions. Much of this literature has been inspired by the seminal work of American linguist George Kingsley Zipf, who found that the frequencies of words use in a linguistic community can be described by a single mathematical formula. The most frequent word in a language occurs twice as much as second most frequent word; the second word occurs twice as frequently as the third, and so on (Zipf, 1949). Zipf's law was confirmed for most human languages. Other scholars have expanded this concept to study the relationship between the use frequency of certain words and their lifespan, i.e. how rapidly they disappear (Lieberman et al., 2007).

In this paper, we apply the concept of inequality and inequality measurements in a new way to study social media. To test our ideas, we will investigate the distribution of social media activity in different parts of Manhattan. We may expect that this activity is distributed unevenly, but what are the exact quantitative characteristics of this inequality? Is social media inequality smaller or larger than other kinds of inequality such as income and education?

## 2.2. Inequality Indexes

What is the correct way to measure inequality? How can we turn our perceptions of inequality such as different levels of wealth or different biodiversity of habitats into numbers? There is no perfect single way to do this. Since Gini published his definition that became known as the "Gini index," a number of different measures have been proposed over the next hundred years, each with its own advantages and particular flaws.

The Gini index is still the most popular inequality measurement, particularly in economics. As we explained above, it takes values between 0 ("complete equality") and 1 ("absolute inequality"). For example, if we look at 2015 World Bank Gini Index of income inequality for countries, we find that it ranges from 0.25 (Sweden) to 0.66 (Seychelles).

While the Gini index (also called Gini coefficient) is the most widely used measure for income inequality, it has also been criticized for its limitations (Litchfield, 1999). The main criticism is that the coefficient is too sensitive to changes in the middle of the distribution, while almost completely neglecting changes at the extremes (Cobham and Sumner, 2013).

Some of the simplest alternative measures of inequality are the 80/20 and 90/10 ratio. These measures use the ratios between different percentiles of a distribution. For example, to calculate the 80/20 inequality measure for the numbers of Instagram images shared in Manhattan Census tracts, we would first count how many images are shared in each tract. Next, we sort all tracts by this count, and find how many images are together in tracts that correspond to 80th and 20th percentile. The 80/20 ratio is simply these two numbers divided.

For 90/10, we consider 90th/10th percentiles rather than 80th/20th. The larger the ratio, the higher the inequality. Contrary to the Gini coefficient, this measure is easy to interpret: how many times larger is the amount of images taken at the tract at the 90th percentile relative to the tract at the 10th percentile. Another advantage of this measurement is that it is not sensitive to outliers, because we are only looking at the differences between two percentiles. However, this also creates a disadvantage because these ratios do not reflect the patters in other parts of the distribution (The World Bank, 2000).

Among many other proposed measures, the Theil index and the Hoover index are most popular. The Theil index measures the distance of how far a given distribution is from a flat distribution, corresponding to a completely egalitarian state. Like with Gini index, higher numbers indicate more inequality. The Hoover index corresponds to the portion of the total income that would have to be redistributed (i.e., taken from the richer half of the population and given to the poorer half) to achieve perfect equality.

These indexes are sometimes reported together, since each has some advantages and disadvantages. They can be easily compared because they share the same idea: the larger ration means a more unequal distribution, while a smaller ratio means a more egalitarian distribution.

Note that like other statistical measures of central tendency and dispersion that were popularized in the early 20th century (mean, median, variance, standard deviation), each of these inequality measures summarizes a set of numbers using a single number. This is both their strength and weakness. Being able to characterize each distribution using a single number allows us to easily compare many distributions. For example, we can compare distributions of income across many countries. At the same time, a single statistics such as Gini coefficient does not capture potentially important differences between the distributions. Using a few different measures together can partly help to overcome this limitation (Jenkins, 1999; De Maio, 2007).

Ecologists comparing species in different habitats use the term "diversity" rather than "inequality." We think that various indexes used to measure biodiversity can also be used for the analysis of social media. An index of diversity measures how many different species are in a habitat, and how many individuals of each species are present. This idea can be also applied to any dataset that contains a number of different categories, with each category being represented by a different number of items.

For example, let's say that we count many different unique hashtags Instagram users assigned to their images shared in many areas. We also count how many times each of the unique hashtags occurs. We can then calculate the diversity of each tract, and compare these diversity measures. In this example, an area of a city such as a Census tract is an equivalent of a biological habitat, each unique hashtag is an equivalent of a particular species, and each occurrence of a particular hashtag is an equivalent to an individual (i.e., a plant or an animal) of the species present in a habitat. Among the diversity indexes used in ecology, one of the most popular measures is called Shannon index, because it uses the same formula originally developed by Claude Shannon to measure entropy (i.e., information content) of a signal (Shannon, 1948).

We can use both inequality and diversity measures to analyze the spatial patterns in social media. For example, if we want to compare the total number of photos shared in every tract in NYC, we can calculate one of the inequality indexes. But if we want to consider not only the number of photos but also their content (which can be quantified by using a number of discrete categories such as indoor, outdoor, people, selfie, etc.), we can apply one of the diversity indexes. This paper only reports our analysis using inequality measures; the analysis using diversity measures will be reported in a future paper.

# 3. Applying social media inequality concept: a case study

## 3.1. Datasets

For our analysis of social media inequality in New York City and its relation to socio-economic inequality, we used the content from one of the most popular social networks – Instagram. It has the strongest geographic identity among all top social media services. While tweets and Facebook posts can also have geo-coordinates and talk about the local events around the user at the moment of posting, Instagram images often directly capture these events. Alternatively, they indicate user's presence and/or behavior in a particular place. Since we are interested in social media as a proxy for physical behavior in space, it makes sense for us to select Instagram for our analysis.  At the same time, because Instagram posts contain an image or a video, date/time metadata, descriptions, and hashtags, they allow us to study spatial inequality using many dimensions. But most importantly, in contrast to other social media services, image-driven Instagram creates an "image of a city" for both locals and visitors. It is therefore most relevant if we want to understand what such collective images contain, and how their patterns are related to both city's architectural structures (for example, presence of tourist landmarks) and socio-economic social structures (for example, the locations of rich/poor areas)

To create a dataset for our study, we collected all publicly shared geo-tagged Instagram images shared in New York City area between February 26 and August 3, 2014. According to other researchers, during that time about 20% of all images on Instagram had geo-locations. (This means that we don't have other images that didn't have geo-location although they were shared in New York).

Because Instagram API limits how many records can be downloaded per hour, we used the largest third party service Gnip to collect the data. Gnip's Geo-collector relies on Instagram API that allows downloading data in a rectangular area centered on a particular geographic coordinate. We stored the data delivered by Gnip into MySQL database using a custom Python script. To download the actual Instagram images, their URLs collected by Gnip were passed to another custom Python script.

Since Instagram did not support downloading large volumes of historical data, we had to download data and images continuously during the period we wanted to cover. A single iMac computer running 24/7 continuously was used for downloading this data. Our preliminary tests showed that even using Gnip that does not have the download limit of Instagram's own API would not allow us to capture data for all NYC boroughs, and therefore we decided to limit our dataset to Manhattan area. Using Gnip, we defined seven overlapping rectangular areas that covered all of Manhattan. Because we use large rectangular areas, we also captured data and parts of Brooklyn, Queens, and the Bronx. The overall dataset we obtained contains 10.5 million images and associated metadata: locations expressed as longitude and latitude, month, date, hour, minute and second when images were shared, user's descriptions and hashtags, and also Instagram username and URL for each image. We have username to calculate statistics of number of images per user and number of unique users per area, and we report these aggregated statistics. However, we don't use any usernames in any figure in this paper.

To be able to research the questions we are interested in, we filtered this dataset in two ways, as described below.

One of our goals was to relate measures of social media inequality and measures of socio-economic inequality. Since the Census only reports aggregated socio-economic indicators for residents, to make our comparisons meaningful we needed to filter out Instagram images shared by visitors. According to 2014 data, there are 1,636,268 people residing in Manhattan (U.S. Cen-sus Bureau, 2014). In the same year, 56.4 million people visited NYC (NYC statistics, 2015). Although we don't know if each of these visitors spent time in Manhattan, if we assume that all of them visited it for at least one day, this makes it 35 visitors for every resident (53.6 million divided by 1.6 million).

The massive number of city tourists mean that if we are interested in geographic patterns of social media inequality - how many images are shared in each part of the city, when they are shared, what are these images, etc. - looking at locals vs. visitors would produce very different results. Visitors have different interests, time availability and routines compared to locals, so any analysis of social media inequality in any major city needs to separate these two groups. Here we can recall again Fischer's project where he compared locations of Flickr images for visitors and locals in over 100 cities, revealing of how dramatically different are the patterns of movement and image capture for the two groups (Fischer, 2010).

Instagram users typically do not report the places where they live, but we can use metadata about the photos they share to probabilistically estimate if a particular person lives in a given city or is only visiting. However, this requires downloading data for all geo-coded images shared by users in any location. For our study, we employed a simpler method that was already used in a number of published papers.

We have used 2014 data from the U.S. Office of Travel and Tourism Industries to calculate the average duration of visitors stay in New York City. The Office report the average duration of stay for visitors for many countries. Averaging these durations, we found that the average visitor stays 10.5 days in New York City. For our study, we decided to use a slightly larger 12-day period. If a user posted all her photos within a single 12-day period out of the total five months of our data collection, we consider this person a "visitor." If a user shared a minimum of two photos within any interval larger than 12 days, we consider this person a "local." Although this very simple method is expected to produce some errors, we felt that they are acceptable given the size of our dataset (10.5M images).[1]

After filtering our 10.5 million images dataset into two groups - one with all the images that were likely shared by visitors and another one with all the images likely shared by locals - we applied a second filter to leave only those images shared in Manhattan. We did this by calculating if the geographic coordinate of an image was inside any of the Census tracts for Manhattan. All images with coordinates outside of Manhattan tracts were then deleted.

The result of these two filtering operations are two final datasets used in the paper. One dataset contains 5,918,408 million images from 366,539 unique Instagram accounts of local residents. (We are aware that this set contains some commercial accounts and also some "super-user" accounts who post much more frequently than other regular users, but for this case study, we did not want to filter out such users, because we are interested in geographic patterns of social media activity rather than individual users.) The second dataset contains 1,524,046 images from 505,345 accounts that most likely belong to visitors.

---

[1] We do recognize that this method is not infallible and probably some of the users whose images were left in our dataset were in fact shared by tourists. Apart from the tourists who stay in NYC

We are also using selected socio-economic indicators from the U.S. Census for all tracts in Manhattan. The particular indicators are from the 2012 American Community Survey (ACS) 5-year estimates which we downloaded using R and ACS.R package (Glenn, 2011). (This was the latest published dataset when we were analyzing the data).

While Census and ACS survey datasets contain hundreds of indicators, we wanted to work with a smaller list that exemplified the types of socio-economic indicators commonly used in social science. Accordingly, we selected the following indicators per tract from the ACS dataset: median income, median rent, percent of people with health insurance, average age, race composition, and average commute time to work.

Census tracts are small and relatively stable subdivisions of counties used by the U.S. Census Bureau. The latest full 2010 census used 287 census tracts for Manhattan. The small size of the tracts allows us to compare selected Census indicators and social media use across all these micro-parts of Manhattan. Since the Census data is aggregated for all people residing in each tract (no individual data is reported), we have similarly aggregated Instagram characteristics for the corresponding tracts. This allows us to compare social media inequality and socio-economic inequality at the tract level.

We also used the 2012 ZIP Code Business Patterns dataset by the U.S. Census Bureau. This dataset specified the number of business in the U.S. geographic areas for different industries and business sizes in terms of number of employees. Because this data is reported at the ZIP code level, which in the case of Manhattan are larger than most tracts, we matched ZIP code data to individual tracts. This data allows us to calculate distributions of numbers of businesses and their sizes on tract level.

It is important to stress again that our study looks at social media inequality *across geographic areas*, rather than between individual Instagram users. While it would be meaningful to calculate inequality measures for users of Instagram in Manhattan, this is not our purpose in the present study. Our concern here is with spatial social media inequality. By aggregating social media sharing for each tract in the city, we obtain *new metrics for comparing parts of city to each other*. They can be used to understand and compare areas along with other urban metrics such as residential density, street connectivity, walkability, and measures of people's perceptions of their environments such as sense of safety, novelty, intricacy, etc. (Ewing and Clemente, 2013).

Formulated as a type of spatial analysis, our study compares the parts of the city that attract more people and generate more content shared on social media networks and thus are "social media rich" with parts of the city that are "social media poor." What are the relationships between such social media rich and social media poor areas? Is social media inequality larger or smaller than economic or social inequality in the same areas? Does social media inequality increase worldwide, similar to how economic inequality has been growing in the recent period?

Which parts of the world have the highest social media inequality and which are the most equal? Although our analysis is focusing on one part of a single megacity, it can be expanded to the scale of a country and world as a whole to consider such questions.

## 3.2. Analyzing social media inequality using volumes and locations of shared content

We start our analysis of social media inequality by looking at a single characteristic – number of images shared in different parts of a city during a particular period. Later in the paper we will also use numbers of hashtags included by users on their shared images.

To calculate the amount of inequality, we can divide a city into a number of equal size parts using a grid and then compare how many images were shared in each grid cell. In the case of complete equality, every part will have exactly the same number of images. In the case of absolute inequality, one part will have all the images and the rest will have none.

Since we will later use selected socio-economic indicators reported by Census per tractors of images in the same tract, it is convenient to use tracts as opposed to an arbitrary rectangular grid. Figure 3 shows 287 Census tracts in Manhattan with colors indicating the relative number of images shared in every tract by local users and hashtags they added to these images.

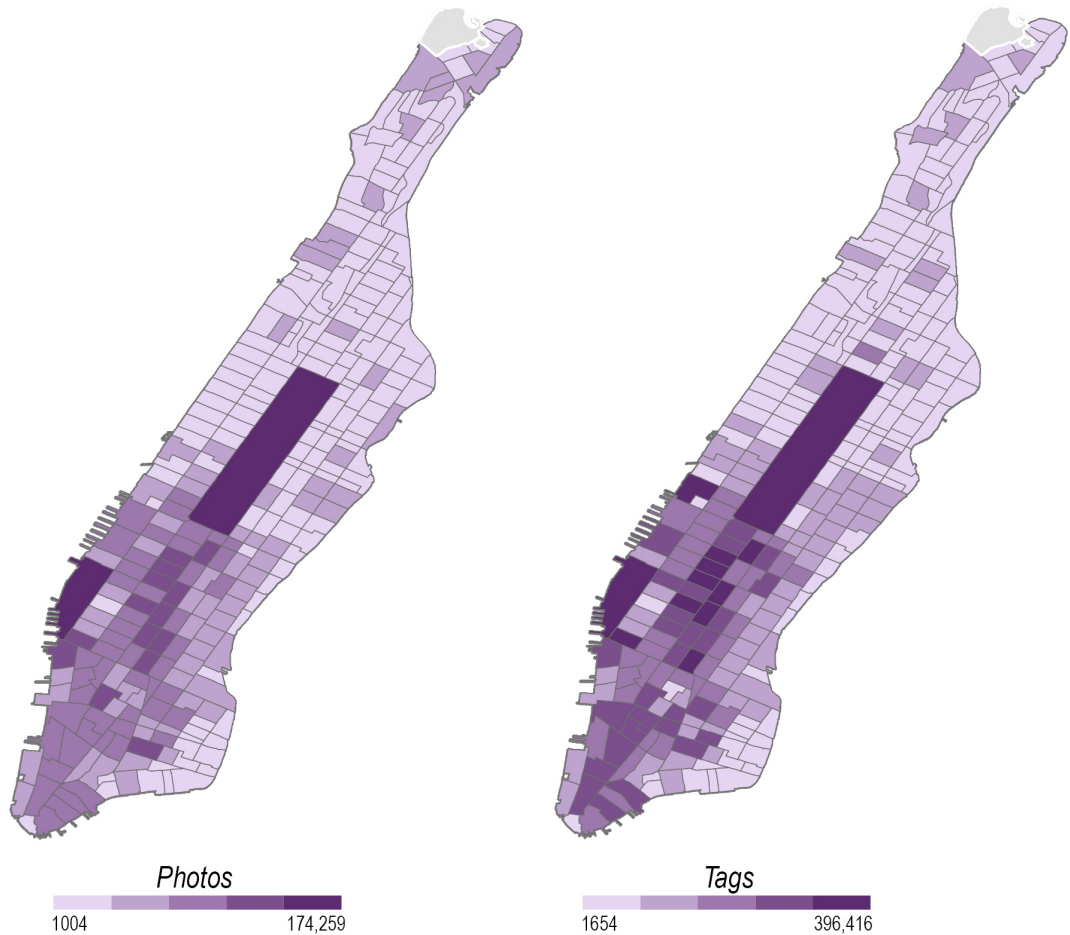

*Photos*

1004    174,259

*Tags*

1654    396,416

**Fig. 3.**

Now that we have aggregated number of images per tract, we can use standard measures for measuring inequality. Since the Gini coefficient is the most popular method for measuring inequality and is used in many fields, we will start by using it to measure inequality of spatial distribution of Instagram images. Confirming what we already noticed in figures 1 and 2, the inequality turns out to be much larger for visitors than for locals: 0.661 and 0.468, respectively. Thus, if we use the Gini measure, visitors' social media inequality is 1.41 times larger than locals' inequality.

The likely explanation is that visitors tend to capture and share images only in particular parts of the city, ignoring many other parts completely. In fact, more than 50% of all images by visitors are shared in only 24 tracts out of all 287 tracts in Manhattan. These 24 tracts cover only 12% of the total area of Manhattan. While the location of images from local users are also dis-tributed non-equally, the amount of inequality is significantly lower: 50% of their images are shared in 53 tracts out of all 287 tracts. These 53 tracts covering approximately 21% of the total Manhattan area.

In general, we may expect that larger geographic areas will have more people living or visiting them and therefore these areas will have more shared images. Given that the sizes of Census tracts vary significantly, with largest tracts 10 times bigger than the smallest tracts, we decided to normalize our data by tract size. In the rest of the paper we use such normalized data.

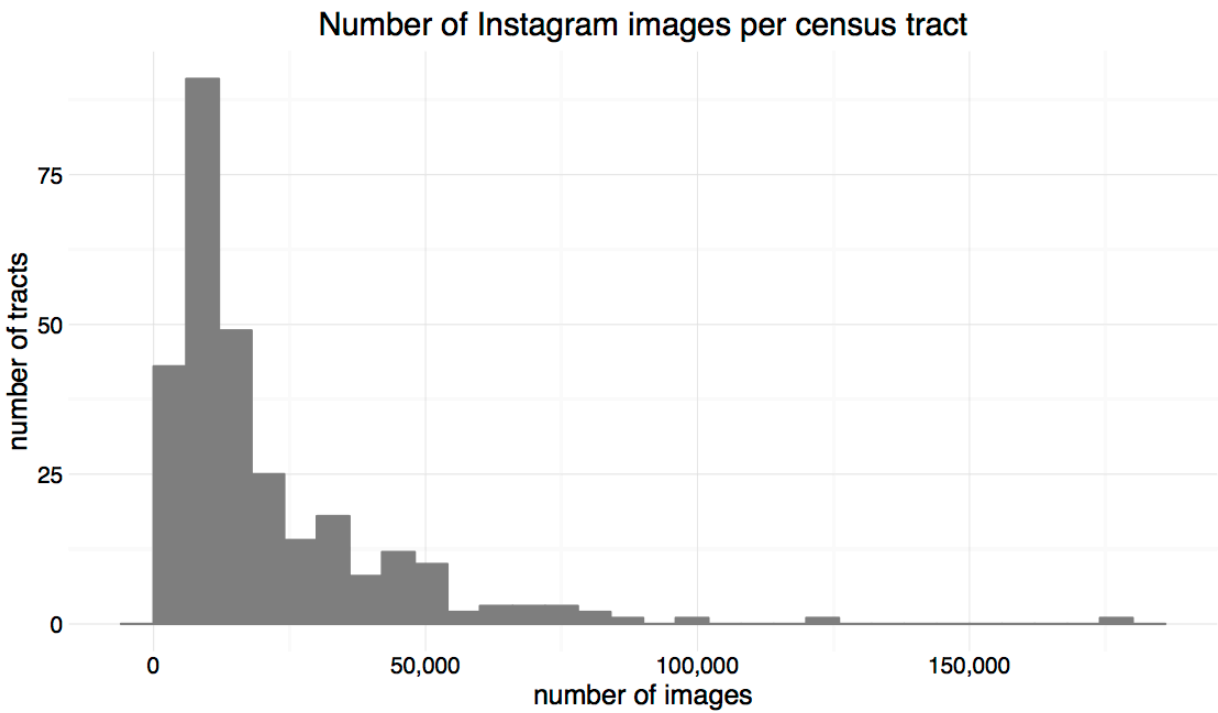

**Fig 4a.**

Figure 4a shows the distribution of numbers of images by locals per square kilometer after the data was normalized. The number of images varies from 2,127 per sq. km to 552,787 per sq. km, and the mean is 106,431. The differences in "social media coverage" between parts of a city are staggering. The ratio between sq. km areas with most (552,787) and least images (2,157) is 256,275, i.e. a quarter of million times!

Fig. 4b and 4c compare numbers of images shared by locals and tourists per tract (also using normalized data). Here again we see how inequality for tourists is much stronger than for locals.

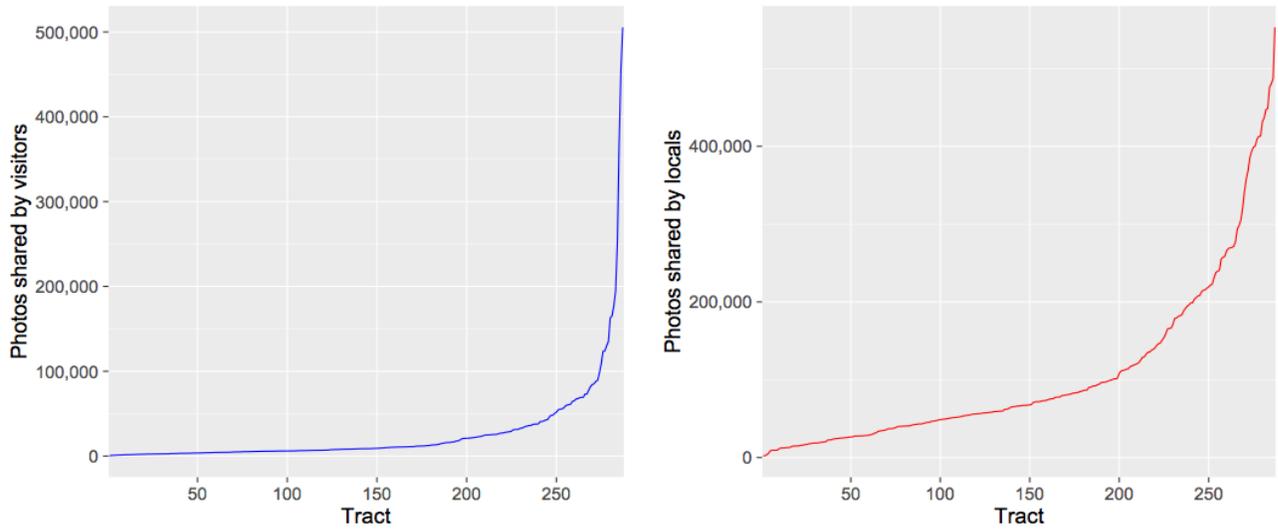

**Fig. 4b and 4c.**

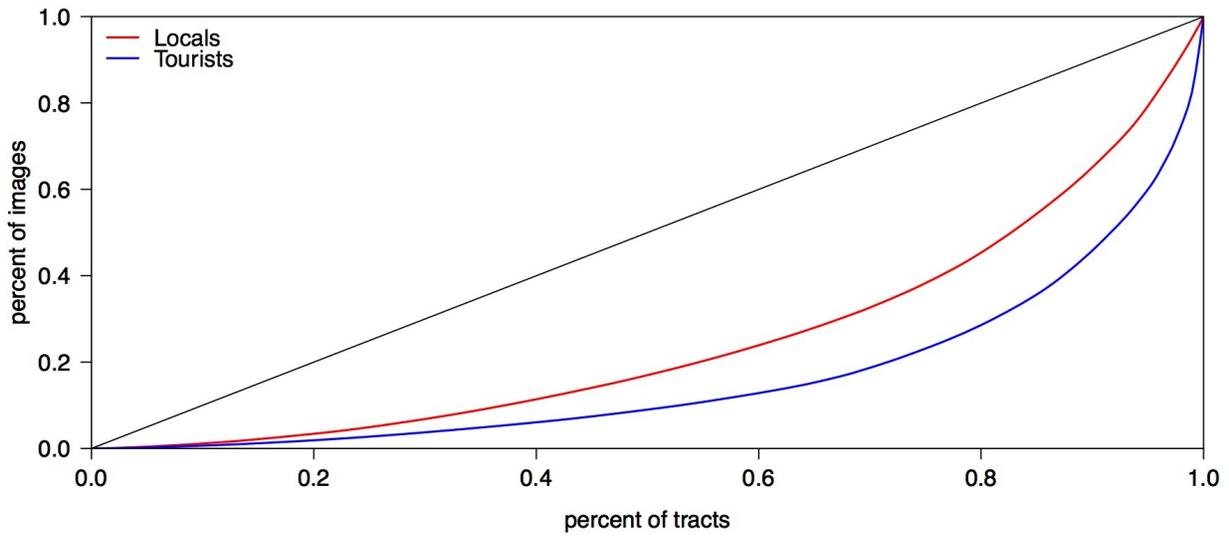

**Fig 5.**

The Gini coefficients for images shared by visitors and locals calculated using normalized numbers are 0.669 and 0.494, respectively. Thus, the true inequality measures are significantly larger than our first calculations that used non-normalized data (0.661 and 0.468).

To put this in context, we can compare our social media Gini coefficients to the Gini coefficients for countries' income. While income inequality and social media inequality are defined and calculated differently, such comparisons are still relevant. If we find a large inequality in any population on any dimension, this is an important characteristic of this population.

Social media inequality of visitors' images in Manhattan (Gini = 0.669) is larger than income inequality of most unequal country in the world (Seychelles where Gini = 0.658). On the other hand, social media shared by locals has a Gini coefficient similar to countries that rank between 25 and 30 in the list of countries by income inequality. These are countries like Costa Rica (0.486), Mexico (0.481) and Ecuador (0.466). (The World Bank, 2015).

Gini coefficient for income in Manhattan according to 2014 estimate was 0.594 (the U.S. Census Bureau, 2014). "New York County (Manhattan Borough)" Census area was the most unequal among all counties in the U.S. Interestingly, this number lies approximately in the middle between social media inequality for Instagram calculated for visitors and locals (0.669 and 0.494, respectively). Note however that income inequality in Manhattan keeps increasing, so Gini coefficient is likely to be larger by the time you are reading this paper.

The Gini index can be visualized using Lorenz curve method. Lorenz curve was introduced by American economist Max Otto Lorenz in 1905 to represent wealth inequality (Lorenz, 1905). The curve represents the cumulative amount of a resource owned by all people in the society or social group. Any particular point on the curve shows how many people (X axis) own a particular share of a resource (Y axis.) In a perfectly equal society, everybody has exactly the same amount of the resource, and Lorenz curve becomes a straight line at a 45-degree angle. But in real-life economies some people will have more of the resource than others. The more unequal the distribution, the larger the deviation of the curve with respect to the 45-degree. Lorenz curve method is used in different fields, so the resource shown does not need to be only income or wealth, and X axis can also represent entities other than individuals.

Figure 5 visualizes Gini inequality measurements for images shared by visitors and locals using Lorenz curves. In contrast to single Gini numbers, Lorenz curves give us more information, showing the patterns in the bottom, middle and top parts of a distribution. In our case, the curves also show that both distributions have similar shapes.

How would our results change if we use alternative inequality measures? Using again normalized data, we compute 80/20 and 90/10 ratio. To recall, the first measure is a ratio between the numbers of images for all tracts in the 80[th] and 20[th] percentile. The second measure uses tracts in the 90[th] and 10[th] percentile. Each measure captures different aspects of the distribution. The first includes 40% of the distribution, so it is more inclusive. The second measure includes only 20%, so it better captures the differences between the extreme ends of the distribution.

80/20 ratio is 7.9 for visitors and 6.0 for locals. This ratio of 6.0 means that the tracts at the 80th percentile contain six times more images than the tracts at the 20th percentile. However, if we only count the extreme tails of the distribution and compute 90/10 ratio, the difference is many times larger. 90/10 the ratio is 25.0 for visitors, and 13.9 for locals. According to this measurement, social media inequality for Instagram in Manhattan shared by visitors is almost two times larger than for locals.

While each of the different measures is useful in its own way, we also get new insights by comparing them. For example, we now know that the ratio between visitors' and locals' inequalities is larger if we use 90/10 method as opposed to 80/20 method (1.80 vs. 1.32, respectively). This indicates that the extremes of the distribution are responsible for most of the difference between visitors and locals. That is, the difference between the number of images shared in the most popular tracts and the least popular tracts is much larger for visitors than among locals.

For visitors, the top 10% of tracts contain 54% of all their images, while the bottom 10% of tracts contain less than 1%! For locals, the top 10% of tracts contains 35% of all images, while the bottom 10% contains just 1% (same as for visitors). This adds a new very important detail to the picture that emerged so far in our analysis. The difference between the two user groups does not lie symmetrically in both ends of the distribution, but only in one (top) end. The least popular tracts contain a similar percent of total images for the two groups; it is among the top tracts that we see a wide disparity.

Another inequality measure that we will use is the Hoover index. In our case, it shows the proportion of all images shared across Manhattan that would have to be redistributed to achieve perfect equality. The larger the number, the more unequal the distribution. For locals the Hoover index is 0.37 and for visitors it is 0.52. And finally, the Theil index measures how far the distribution is from being perfectly equal. Again, the larger the index, the higher the inequality. Theil index for visitors is 0.93 and for locals is 0.41. Table 1 compares all five inequality indexes that we used and their values (for normalized data). The right most column compares the measures for visitors and locals by dividing the corresponding inequality measurements for each method.

| Inequality measurement | Visitors | Locals | Ratio between visitors vs. locals measurements |
|---|---|---|---|
| Gini index | 0.669 | 0.494 | 1.354 |
| 80/20 ratio | 7.9 | 6.0 | 1.316 |
| 90/10 ratio | 25.0 | 13.9 | 1.798 |
| Hoover index | 0.52 | 0.37 | 1.40 |
| Theil index | 0.93 | 0.41 | 2.258 |

**Table 1:** Five inequality indexes capture social media inequality for visitors and locals in Manhattan (using locations of Instagram images shared by the two groups in Manhattan.)

To further reflect on the measurements provided by these five indexes, we come back to figure 1. While only some city areas are "activated" by social media shared by tourists (left), social media shared by locals (right) covers most of the city, although some prominent empty parts still remain. Such visualizations give us much more information than couple of numbers calculated using any inequality index or ratio. But if we try to quantify our perceptions using only the visualizations (for example, how many more images were shared around Times Square by visitors vs. locals?), this is not easy. The indexes and ratios have the advantage of precision. They also allow us to quantitatively compare inequality among thousand of areas, which is particularly important if we want to compare global patterns.

But this precision also has its limits. Each inequality measure captures only some property of the distribution but not all its details. What this means, in our view, is that both visualizations and measurements should be used together in studying social media inequality. While such calculations were very time consuming to create at the time when Gini defined the original inequality measure, nowadays a computer does all the work and therefore they should be used as equally important research instruments.

## 3.3. Adding time dimension

In the previous section we analyzed spatial social media inequality for Instagram images shared in Manhattan. To do this, we aggregated locations of millions of images shared over 287 tracts and then compared differences in the volume of images between these tracts. But we have not yet taken advantage of the key difference of social media data from typical 20th century social data – its temporal granularity and density.

Each image shared on Instagram has a time stamp specifying the date, hour, minute, and second when the image was shared. Therefore, we can calculate how many images were shared in a given time interval. Therefore, similar to how we did with space, we can apply inequality measures to compare differences in popularity of time intervals. For example, we can combine data for images shared by locals in Manhattan over five months to calculate average numbers shared per day of a week, and then compare daily volumes. If people share the same number of images every day of the week, the temporal inequality across days for an average week will be 0. If the numbers differ very substantially between days, the "temporal inequality" will be close to 1. Using this logic, we can ask all kinds of questions about temporal patterns. Is weekly inequality bigger for visitors or locals? What are the inequality patters for days of the week, hours of the days, seasons of the year, and so on.

Because the types and the "volume" of human activities change significantly between hours of a day or day of the week – being at work, being at home, sleeping, being active, being with family or friends, commuting, and so on – we need to consider the temporal dimension of social media. But most importantly, the availability of both spatial and temporal metadata for social media content allows us to conceptualize and study cities in new ways. Rather than thinking of social media inequality as a characteristic of a geographic area, as we did in the previous section, we can view it as a dynamic spatiotemporal variable. From this perspective, a city appears not as a static collection of buildings, their residents, firms producing products, and public places but as the aggregations of individuals that follow periodic rhythms in space and in time.

The first temporal analysis of Instagram city patterns was presented in Hochman and Schwartz (2012). In *Phototrails* project and accompanying paper Hochman and Manovich extended this work by analyzing spatial and temporal patterns for 13 global cities using 2.3 million Instagram images shared over a few months (Hochman and Manovich, 2013). Fig. 6 and 7 are two of the visualizations from this paper. Fig. 6 shows temporal patterns in Instagram images shared in parts of New York City (top) and Tokyo (bottom) over continuous time periods. 50,000 images from each city area are visualized in the order they were shared, top to bottom and left to right. Fig. 6a is New York, and Fig 6b is Tokyo.

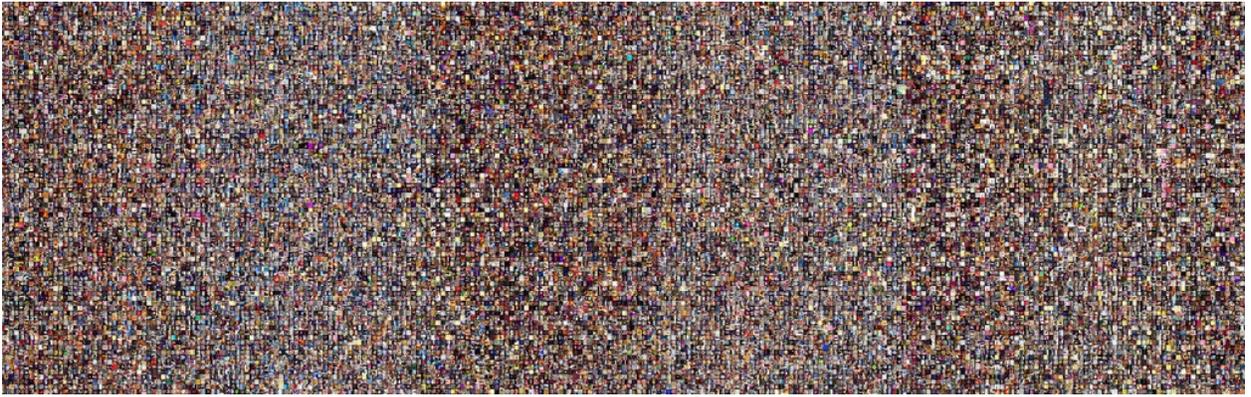

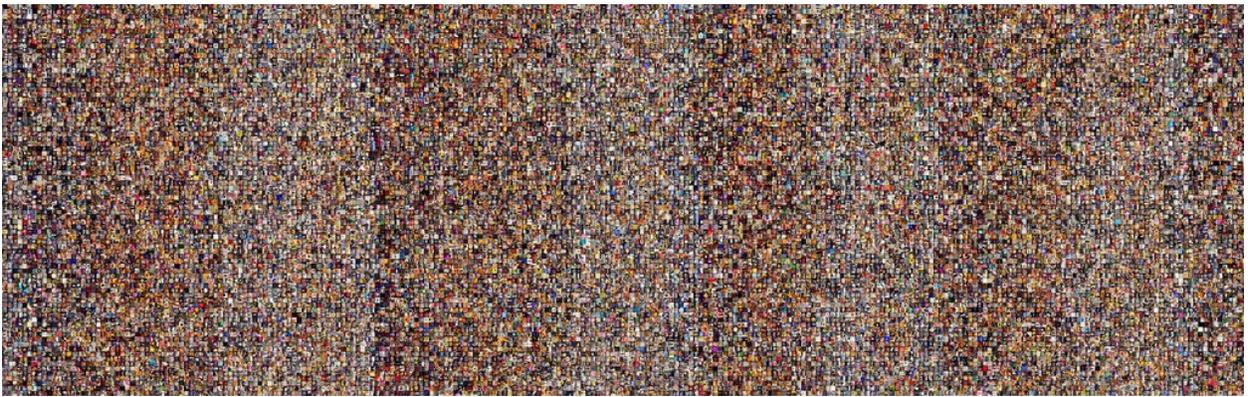

**Fig. 6a and 6b**.

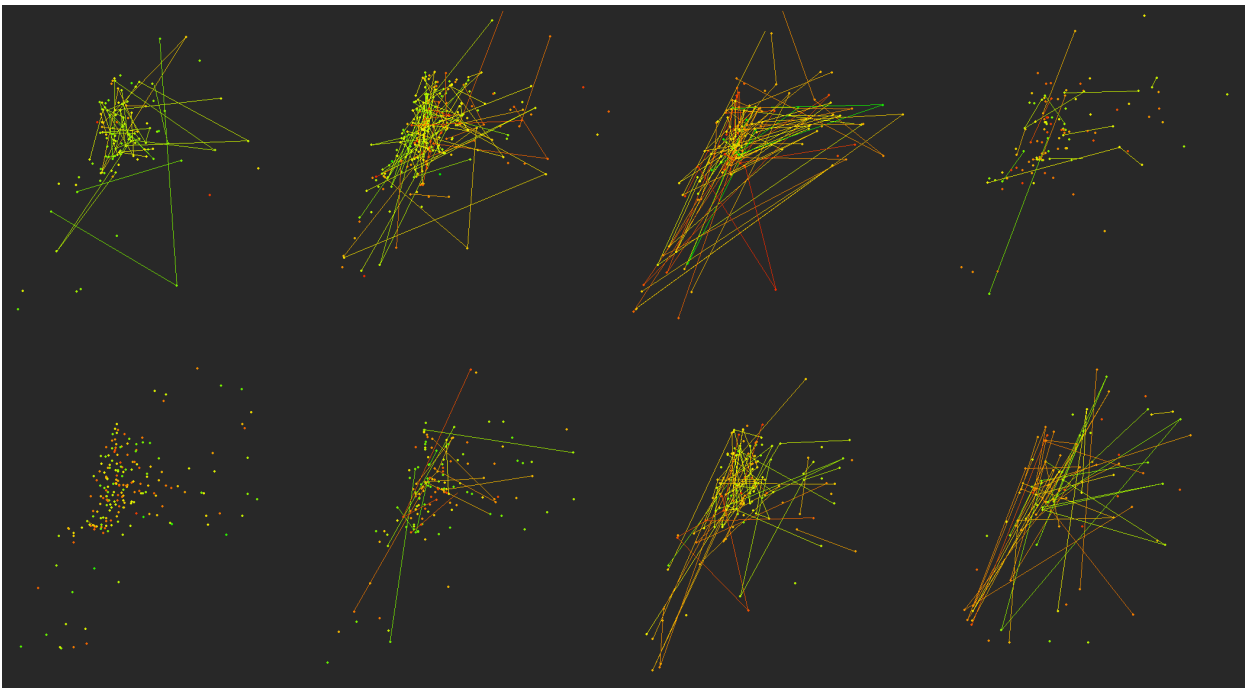

**Fig. 7.**

We can see repeating day to light patterns in brightness - lighter during the day, and darker at night. But each particular 24-hour interval in every city is also unique. Some days on Instagram are longer (more images are shared), and some are shorter. The colors are also not exactly the same in each period. The uncoordinated images shared by thousands of people at the same time inside city area come together to form a "city symphony," with each "instrument" adding its own unique signature. As we can see, the temporal image of a city on Instagram alternates between repetition and variation, predictability and unexpected events, following routines and breaking them. (For the purpose of comparison between many city areas, many cities and many periods, we can disregard these variations and create statistical models that account for the regular part. But as representations of complexity of city life, such visualizations consisting from the actual shared images have their advantages, since they show both the regular and the irregular.)

Fig. 7 shows patterns in time and space together on the level of individuals. The original matrix plot compares 289 Instagram users in Tel Aviv that uploaded most Instagram images with geo-locations during three months in spring 2012. Here we only show a few plots. Each plot shows locations of photos shared by a particular person during this period. The green to red color gradient indicates the time when an image was shared (green — morning, yellow — afternoon, red — evening). A line is drawn between two dots if corresponding photos were shared within the same hour. If we consider total social media content as a type of resource produced by people in the city, quantifying inequality allows us to understand how this resource is distributed spatially and temporally. We may expect that every city will have its own distinct signature of spatial-temporal social media inequality. These signatures reflect where people who share content on a particular social media service or services spend their time, including the waves of commuters traveling daily for work, locals going to other areas for leisure activities, visitors shopping and sightseeing, and so on. Many areas get "activated" during different days of the week and hours of the day. Each can also have different types of users being more or less active at different times.

As we already noted, analyzing and visualizing these patterns moves us away from the image of a city as a static map of physical structures that change very infrequently. Instead, we get a multi-dimensional "volume" that reflects where people are and what they do every hour. Three dimensions of such volume would correspond to space; another dimension would correspond to time; others can indicate types of users; still others would code different kinds of social media characteristics such as volumes of messages, their content uniqueness, etc.

Figure 8 shows three such slices of the "volume" of Manhattan - hourly proportions of images shared by locals and by visitors in selected neighborhoods. To show patterns in a 24 hour cycle, we aggregated all data we have for five months. The graphs reveal important differences in Instagram sharing between neighborhoods and between locals and visitors. (Because the volume of images shared by visitors in the neighborhoods above 110[th] street is quite low, these patterns are not as reliable, so we are not including this graph).

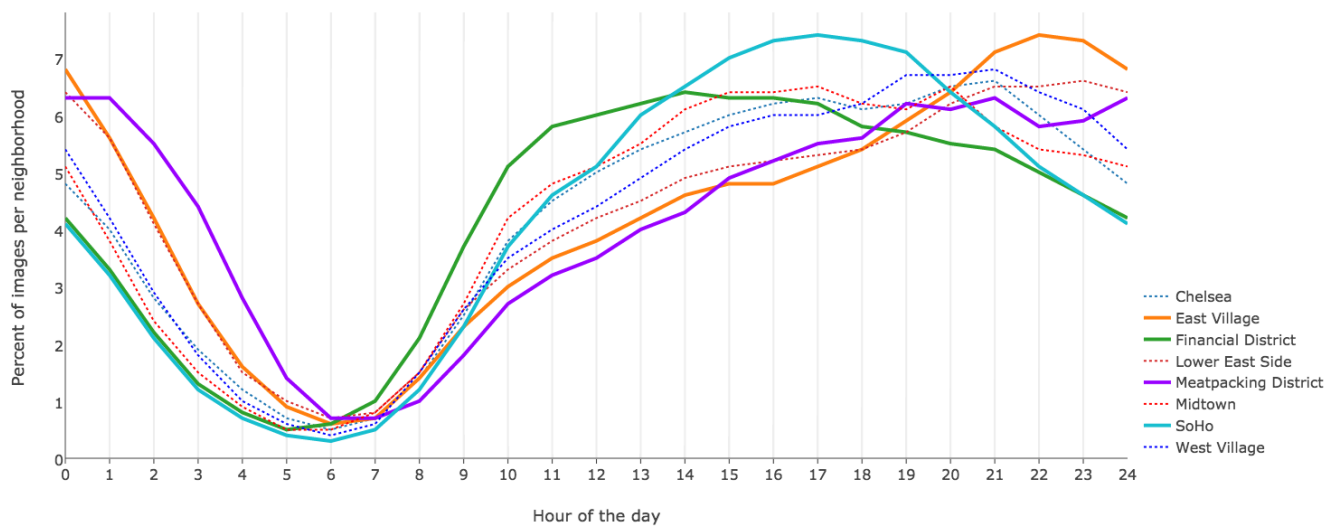

Locals in selected Manhattan neighborhoods - 1

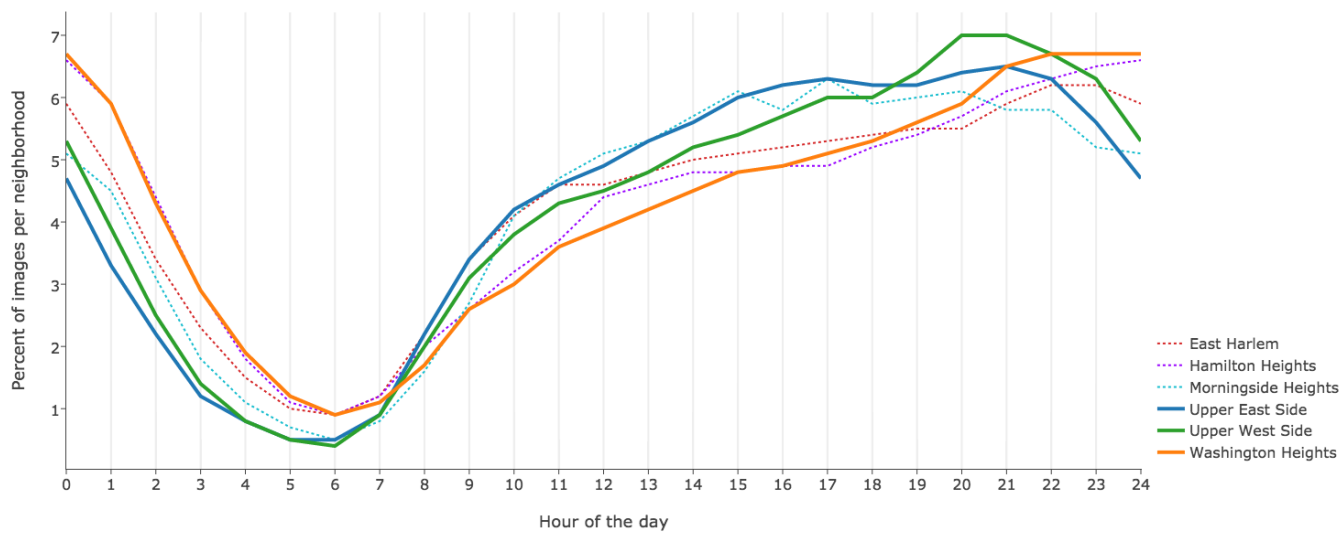

Locals in selected Manhattan neighborhoods - 2

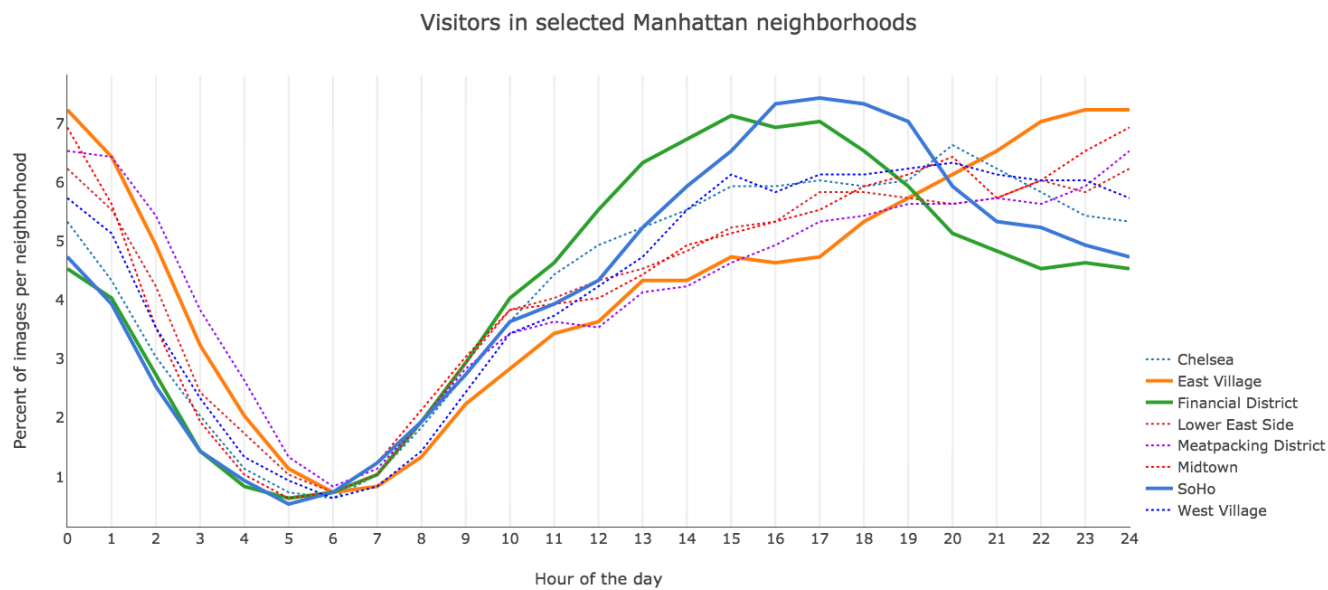

**Fig. 8a, 8b and 8c.**

For locals, we see a clear separation between work areas (Financial District), shopping areas (SoHo) and nightlife areas (East Village, Lower East Side, Meatpacking District). For visitors, these patterns are similar but even more clearly pronounced (for example, volume in East Village keeps increasing after 9 pm). One exception is Midtown: here visitors' activity keeps increasing late in the evening. The possible reason is that this area has the biggest concentration of hotels in the city and also key attractions such as Times Square.

The patterns in the neighborhoods located above 59th street are different. Overall, the activity keeps increasing and peaks later in the evening. And in the areas with lower average income (East Harlem, Morningside Heights, and Washington Heights), the activity keeps increasing until midnight. We will come back to these differences in the next section where we will offer a possible explanation.

Many other slices can be equally revealing. Analyzing data in every slice will produce a different social media inequality measurement. This suggests that to better characterize social media activity in a city, we need to measure the "inequality of inequalities" – for example, the distribution of inequality indexes for each day in a year, or the distribution of spatial inequalities at different spatial scales, and so on. Construction of such "super-index" can be the interesting subject for future research.

We should also mention another important consequence of considering both time and space together in social media analysis. So far we looked at both time units such as hours and spatial "semantic" units such as neighborhoods as fixed entities. However, the continuously changing patterns of sharing create clusters in time and in space. In one time intervals some neighborhoods may have similar patterns, thus forming a single larger cluster. In another time interval we may see smaller parts of different neighborhoods having the same pattern, but not a neighborhood as a whole. This does not mean that the boundaries of "neighborhoods" (or another type of spatial unit) are completely irrelevant for a "social media city." Rather, social media shares are likely to construct their own map of divisions that change periodically over time. Sometimes they may overlap with neighborhood divisions, and other times they may have little in common with them. The same holds for time. A 24-hour cycle may get divided into a few periods depending on volumes of shared images, gradual or rapid increase or decrease, or other patterns. We can see such patterns for some of Manhattan neighborhoods in Fig. 8. (Manovich, 2014 presents the analysis of the temporal patterns in central areas of six global cities.)

If we aggregate the data without separating neighborhoods or tracts, this will hide important differences seen in Fig. 8. However, despite their differences, temporal patterns in all shown neighborhoods have a similar shape - the least amount of activity is around 5am, then gradual rise that continues until late afternoon or into the late evening, and then decrease in activity after midnight. This suggests that aggregating data for Manhattan as a whole is also meaningful and can give us additional insights. In the following we present such analysis.

So far we included all local users as long as they shared at least two images during our five month period. This made sense since we were comparing numbers of shared images across all Census tracts or selected neighborhoods. Together, all images shared on Instagram in Manhattan contribute to the collective "portrait of a city" accessible to any user of Instagram around the world, so we had no need to exclude any of them.

To compare temporal patterns, we decided to only look at local users who have been using Instagram systematically. To do this, we created a new dataset of "super locals" by only users who shared at least one image within each of five months. If the original locals dataset has 5,918,408 images shared from 366,539 accounts, "super local" dataset has 527,401 images shared from 5360 accounts. However, this smaller set of users have been sharing systematically over a long period of time, so we felt that their temporal patterns are more reliable.

Fig. 9 compares temporal patterns for these super locals and for tourists. The first graph shows volumes of images shared per hour in a 24-hour cycle. The second graph shows volumes of images shared per day of a week. The days are arranged in the following order: Sunday (1), Monday (2) ... Saturday (7). As we can see, super locals become active earlier than visitors. And their activity level also starts decreasing already around 6 pm, while tourists continue to be active until midnight. Both of these patterns make sense.

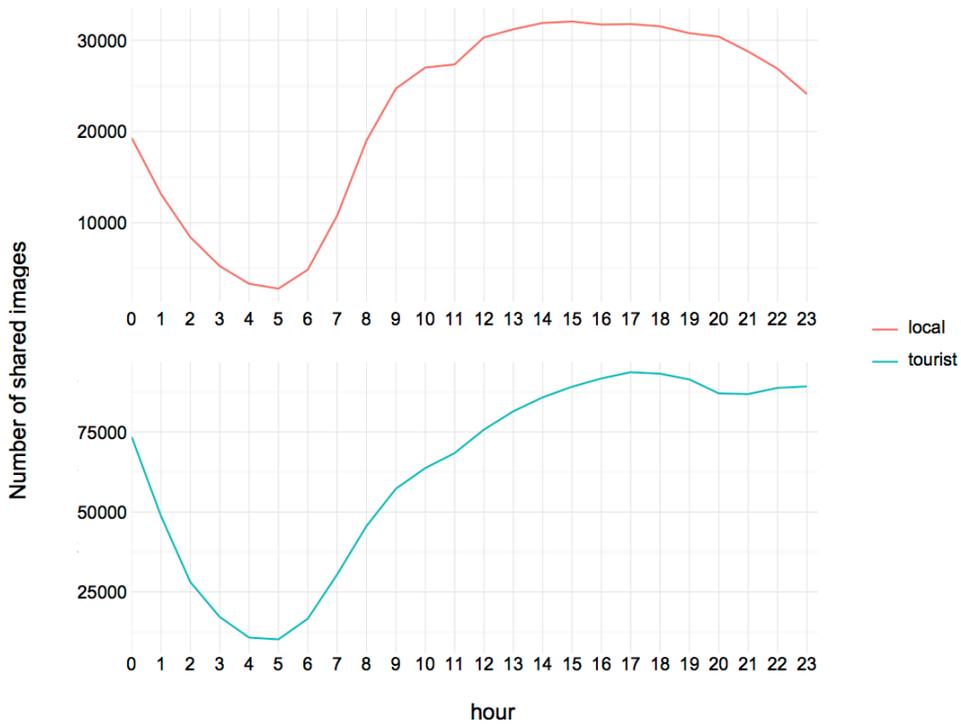

**Fig. 9a.**

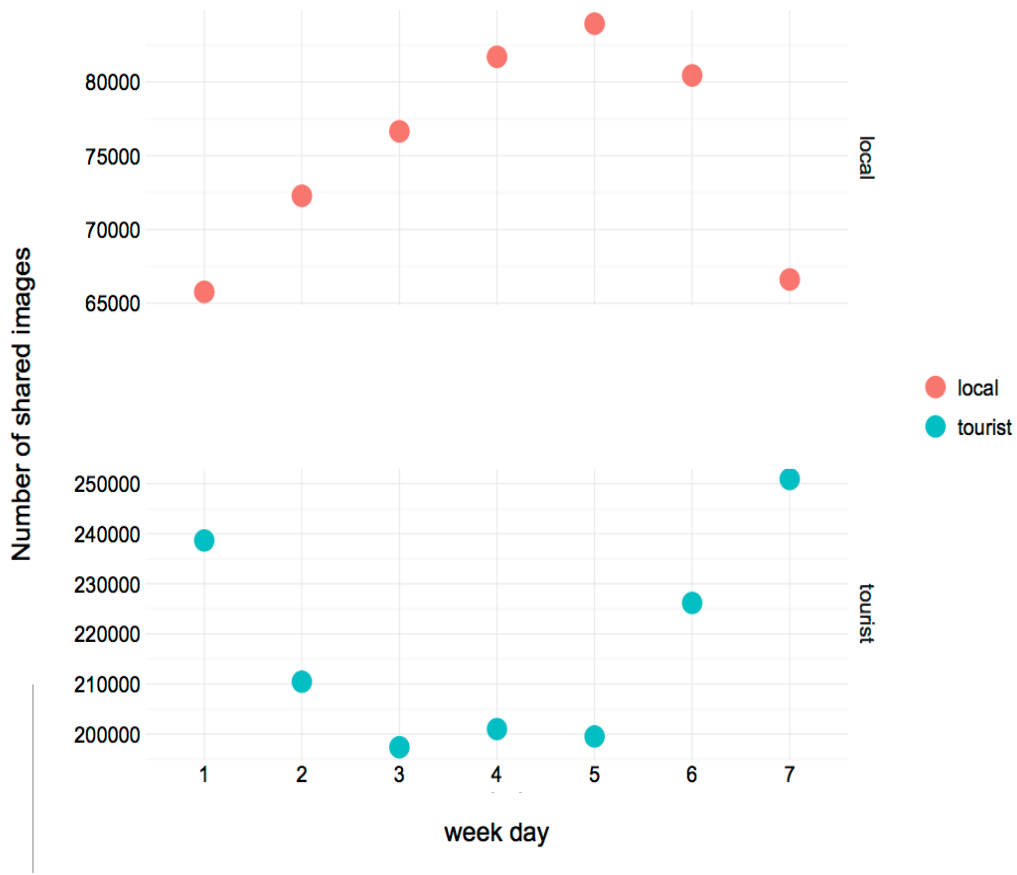

**Fig. 9b.**

The patterns for days of the week of the two types of users are almost mirror reflections of each other. Super locals' activity keeps gradually increasing from Monday through Thursday, and then it rapidly falls off. Tourists' activity is low from Tuesday through Thursday, but high from Friday through Monday. The possible explanation for this pattern is that most visitors are only coming to New York during the weekends. This makes sense if these are visitors that live nearby (New Jersey, Long Island City, etc.) so they only came to Manhattan for one weekend day, or perhaps the whole weekend. Indeed, out of all 56.5 million NYC visitors in 2014, 44.5 million were domestic visitors (NYC statistics, 2014).

(We have also plotted hour and day of the week data for the complete locals dataset. The 24-hour cycle pattern was almost exactly the same as for super locals. The days of a week pattern was also almost the same, but the differences in volume between individual days were much smaller.)

Fig. 9 demonstrates that social media (here, Instagram) indeed captures some types of human behavior quite well, and therefore can be used as indirect, but reliable signal for studying the society - or more precisely, the "society" of people with particular demographic characteristics who are active on particular social networks.

What was this "Instagram society" in Manhattan in 2014? Our "locals" dataset contains 366,539 unique usernames, and as far as were able to determine, almost all of them belong to individuals. Dividing 366,539 by 1,636,268 (the size of Manhattan population in 2014), we get 22.4%. In other words, one in five Manhattan residents shared images on Instagram.

Obviously, this result is not precise: some of the users are companies, and many users could have just tried Instagram once and did not post after that. However, given that we downloaded only approximately 20% of all shared images in Manhattan (because the other 80% did not have public geo coordinates), the actual number of active individual Instagram users in Manhattan could have been even bigger. It is also relevant to note that in 2014, the number of active Instagram users in the U.S. was 62.4 million, and USA population was 318.4 million. Thus, percentage of individuals in the U.S. using Instagram at that time was 19.5%. This supports our estimate of 22.9% Manhattan residents using Instagram in the same year.

Let's summarize our analysis so far. If we consider total social media content as a type of resource produced by people in the city, quantifying inequality allows us to understand how this resource is distributed spatially and temporally. We may expect that every city will have its own distinct signature of spatial-temporal social media inequality. These signatures reflect where people who share content on a particular social media service or services spend their time, including the waves of commuters traveling daily for work, locals going to other areas for leisure activities, visitors shopping and sightseeing, and so on. Many areas get "activated" during different days of the week and hours of the day. Each can also have different types of users being more or less active at different times.

As we already noted, analyzing and visualizing these patterns moves us away from the image of a city as a static map of physical structures that change very infrequently. Instead, we get a multi-dimensional "volume" that reflects where people are and what they do every hour. Three dimensions of such volume would correspond to space; another dimension would correspond to time; others can indicate types of users; still others would code different kinds of social media characteristics such as volumes of messages, their content uniqueness, etc.

# 4. Comparing Social Media Inequality and Socio-Economic Inequality

An unequal distribution of residents on income, ethnicity, and other economic, social and cultural dimensions is a universal feature of all cities in which people have control over where they live (Cheshire, 2009). That is, in any city where people are allowed to live wherever they want, we see a differentiation on some of these dimensions. This has been widely studied by urbanists for a long time, and there is vast literature on residential segregation and inequality within cities.[2]

As we saw, spatial distribution of Instagram images shared by Manhattan residents is also highly uneven. But how is social media inequality related to socio-economic inequality? For example, in a place like Manhattan, is social media inequality smaller, bigger or similar to the inequality of various socio-economic characteristics? In this section we look at this question using selected economic and social indicators for Census tracts on the city and our data about volume of Instagram images shared by locals in the same tracts. The indicators come from American Community Survey 2012 estimates (American Community Survey, 2012). ACS is the yearly estimate published by U.S. Census Bureau based on the responses of a sample of U.S. residents.

We downloaded ACS data using R *acs* package. We then considered a number of socio-economic indicators available per tract: number of households surveyed, median age, median household income, median rent, total population, race, employment status, time of commute to work, educational attainment, health insurance coverage, and Gini coefficient for income. Because most of these indicators are correlated, we decided to use a smaller subset: median household income, median rent, and unemployment rate.

In order to analyze social media inequality together with socio-economic data, we will first look at the differences between numbers of images shared during day and night periods. To do this, we divide the data into two parts. The first part that we call "daytime" contains all images shared between 7:00am and 6:59pm. The second part that we call "nighttime" contains all the images shared in the remaining period, i.e. between 7:00pm and 6:59am.

Next, we rank each tract in two ways. The first rank is determined by numbers of images shared during the daytime. The tract which has the most images shared between 7:00am and 6:59 pm for all days in our five-month period as rank 1; the tract with the second largest number of images has rank 2, and so on. The second rank is determined by numbers of images shared during the daytime, and it is computed in an analogous way.

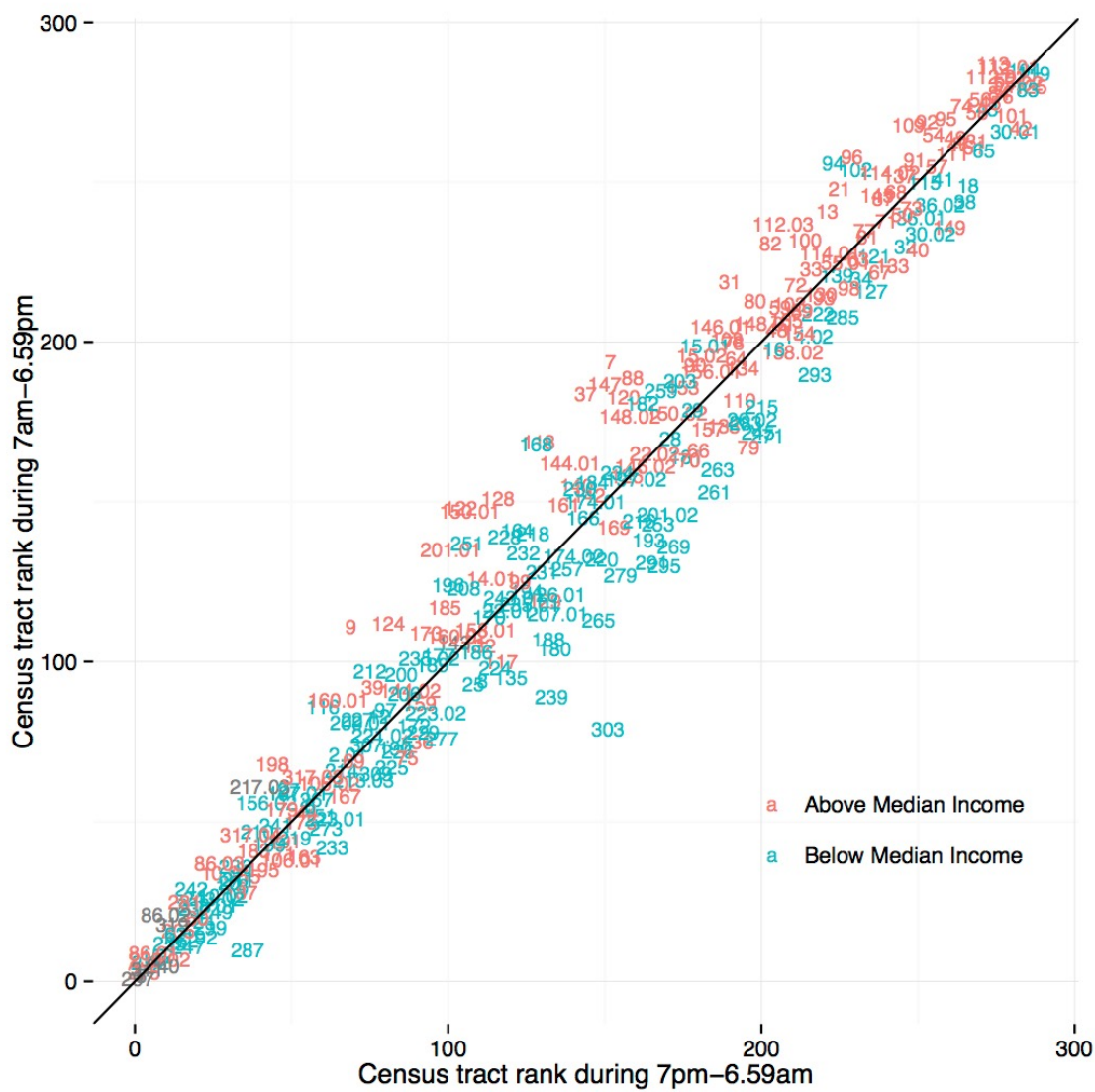

**Fig. 10.**

Figure 10 is the scatter plot of all 287 tracts using these ranks. Each tract is shown as a point; the nighttime rank determines its position on X axis, and the daytime rank determines its position on Y axis. We also use color to code median household incomes (one of the Census indicators) for the tracts. Blue-green indicates that the median household income of that tract is below the median household income for all of Manhattan ($74,693). Red color indicates that median income of the tract is above the overall median income. The number next to each point indicates the number of the corresponding tract used by the Census to identify it.

This plot allows us to see several things. First of all, we see which tracts are the most and least popular (i.e. have the most and least images shared), both overall and also during daytime and nighttime separately. These most popular tracts are in the upper-right corner of the plot. Among the 33% most popular tracts, 72% of them have a median household income higher than the overall median income in Manhattan. Thus, more popular tracts tend to be tracts with more affluent residents. Conversely, among the 33% least popular tracts, 56% have a median income lower than the overall median income in Manhattan.

If people's Instagram activity across the city was exactly similar during daytime and nighttime, we should expect that every tract would have the same rank for both periods. The most popular tracts during the day would also be the most popular during the night; and the least popular tracts during the day would be among the least popular during the night. All points in our plot then would lie on the 45-degree angle straight line. But if daytime and nighttime periods were completely unrelated, we would see points in every part of the graph, without any clustering along the 45-degree angle straight line.

As we see in Figure 10, the points for most tracts in Manhattan indeed lie close to this line. This means that the overall popularity of these tracts is similar between day and nighttime. However, we also see that some points lie at significant distances from the 45-degree line. What can we learn about the temporal dynamics of Instagram sharing in Manhattan and the varying patterns of inequality by examining the tracts corresponding to these points?

First we will consider the points below and to the right of the 45-degree line. They correspond to tracts that are more popular during nighttime than daytime. Checking the locations of these tracts, we find that they are mostly located in the northern part of Manhattan, representing parts of neighborhoods such as Inwood[3] and Washington Heights[4]. In fact, the eight tracts with largest differences between their daytime rank and nighttime rank all belong to these two neighborhoods.

---

[3] Inwood is on the northern tip of Manhattan: north of Dyckman St, between the Hudson River and the Harlem River.

But more interestingly we also find that there is an overall relation between nighttime and daytime popularity and the median household income of tracts. Tracts with median income below the Manhattan average (i.e. median income calculated for all tracts) mostly lie below the 45-degree line, indicating that they are relatively more popular during the night than daytime. In fact, 62% of below-average tracts are below the 45-degree line. For many of these tracts, these differences are large. Specifically, 34 out of the 40 tracts with largest differences between nighttime rank and daytime rank have a median income below Manhattan average.

We will now examine the points that lie above and to the left of the 45-degree line. They indicate tracts that are more popular during daytime than nighttime. Unlike the points below and to the right of the 45-degree line, these points are more spread out geographically, covering Midtown East, the Upper East Side and Wall Street. All these areas contain more affluent tracts. In fact, 57% of tracts with median household income above the overall average of Manhattan are more popular during the day than night.

These patterns can be explained if we accept the following three assumptions:

1) Most people who commute to work in Manhattan do so during the daytime, and that the people commuting to work during the day greatly outnumbers that of people who commute to work at night. "Work" here refers not only to 9 to 5 office shifts, but also meetings, errands, and other work-related activities. People come to Manhattan at some time during the morning or afternoon, do some work and then go back to their residences in the evening.

2) Many people who live in less-affluent tracts work during the day in more-affluent tracts where lots of businesses are located.

3) In Manhattan majority of daytime jobs are in the parts of the city which also happen to have higher median incomes. Specifically, this is the part of the city from the lower south corner to around 100th street.

Our support for this point comes from three different sources. First, we looked at the 2012 ZIP Code Business Patterns dataset by the U.S. Census Bureau. This dataset provides data on the number of business establishments per ZIP code. The largest concentration of businesses in Manhattan is in Midtown and parts of Downtown. On the other hand, the part of Manhattan north of Central Park has much fewer businesses. Secondly, we also base our assumption on Kaufman et al. 2014 who ranked all neighborhoods in NYC according to the number of jobs accessible from the neighborhoods by transit within 60 minutes. They find a strong connection between household income and job accessibility among neighborhoods. The 22 neighborhoods with most jobs accessible within a 60 min radius are all in Manhattan, and have an average income of $108k, which is much higher than the average household income in the city. On the contrary, neighborhoods with the lowest household incomes have significantly fewer job opportunities. Finally, visual evidence from Manduca (2015) also supports this point. Manduca made a map with one dot for every job in the U.S. using the counts of jobs by block from 2010 Census data. This map shows that the majority of jobs in NYC are in Midtown and parts of Downtown, while the northern part of Manhattan seems to have a much lower density of jobs.

---

[4] Washington Heights is bordered by 155th Street to the south, Inwood to the north along Dyckman Avenue, the Hudson River to the west, and the Harlem River and Coogan's Bluff to the east.

If we make these three assumptions, we can easily explain the patterns we observed in Figure 10. The residents of less prosperous areas (such as parts of Manhattan above 100th street) go to work in more prosperous parts of the city (especially Midtown Manhattan) during the day. This is where they share images on Instagram during the day, so their shares get added to these areas. Consequently, since they are absent from their home areas during these working hours, the volumes of images in these areas during daytime is small. In the evening, they return to their areas of residence, and this is why these less prosperous areas have higher volume of Instagram shares at evening and night hours.

Our analysis suggests that differences in social media activity among parts of a city are to a large extent driven by commuting patterns. During weekdays the residents of less prosperous areas (such as parts of Manhattan above 100th street) work in more prosperous parts of the city - areas below 100th street, and particularly in Midtown. This is where they share images on Instagram during the day, so their shares get added to these areas. Since they are absent from their home areas during these working hours, the volumes of images in these areas during daytime is small. In the evening, they return to their areas of residence, and this is why these less prosperous areas have higher volume of Instagram shares at evening and night hours.

Note that according to Pew Internet 2015 data, people in U.S. who have higher education levels and household income are more likely to use social media, but the differences they report are not very big, so they can't account for much larger differences we see in Manhattan in our data. I.e., social media differences between areas are much stronger than the differences in income. Taking into account commuting patterns, we were able to solve this mystery.

Note also that the areas of Manhattan below 100th street with most businesses are also the ones that are the most popular among visitors. Thus, we have the effect of *double amplification* – social media contributions by affluent residents of these areas get amplified with contributions of people who travel there for work and also by contributions from city visitors. This amplification may be the key reason why spatial social media inequality we calculated for Manhattan using Gini coefficient is so high. One part of the city gets images from three groups (residents, commuters, and visitors), while the other part gets the potential images "subtracted" (these are images that would be shared if the residents in this part did not commute).

We may expect that in other geographical areas around the world different relations between places of residence and work, income distributions, and tourist areas lead to different spatial and temporal patterns of sharing. For example, in many European cities, small historical centers are popular with visitors but most businesses employing lots of people are located outside these centers.

Another interesting issue is the effects of changing patterns of work - especially in creative and software industries. Many cities now act as distributed workspaces where designers, programmers, bloggers, and other culture industry professionals work from cafes close to where they live, or co-working places that now number in hundred in NYC. Note that this demographic is also likely to be most active on social networks such as Instagram in many parts of the word.

Finally, consider another issue which is changing many cities worldwide today – gentrification. An area which previously only had less affluent residents, who may be commuting to work during daytime in other parts of the city, may now have a growing proportion of creative class workers and other freelancers who also stay there during the day to work from homes or cafes. Some parts of Manhattan above 100th Street have been undergoing gentrification for a while now. Although our dataset only covers five months and therefore does not allow us to qualitatively analyze the effects of this gentrification on Instagram sharing, the elevated volumes of images in certain areas described as being gentrified suggests that the two are related.

As we explained in the Introduction, in this study we do not directly compare social media inequality and socio-economic inequality at the individual level. That is, we don't try to look at correlations between the volume of Instagram images and income or other socio-economic indicators for every tract. While the Census only counts individuals who have their residence in a tract area, the shared social media images in each tract can also come from daytime commuters, visitors, and other non-residents. At the same time, in many tracts their residents are not there during the work hours, so they can't contribute to this tract's social media presence. If we wanted to directly compare the Census indicators and social media activity in every tract, we would need to obtain Census data for every individual resident, identify if this resident has Instagram account and get the account name. This information is not reported by either the Census or Instagram for privacy reasons.

However, it is meaningful to calculate inequality measures for social media shared by locals and compare them with inequality measures for selected socio-economic variables across all tracts taken together. For our final analysis, we will compare inequality for volume of shared Instagram images and three socio-economic indicators for Manhattan: median household income, median rent, and unemployment rate. The Gini coefficients are 0.32 (median income), 0.22 (median rent), 0.35 (unemployment rate), and 0.49 (numbers of Instagram images shared by local residents). (We calculate first two numbers using published Census indicators per tract.) Figure 11 shows Gini measures for these variables using Lorenz curves.

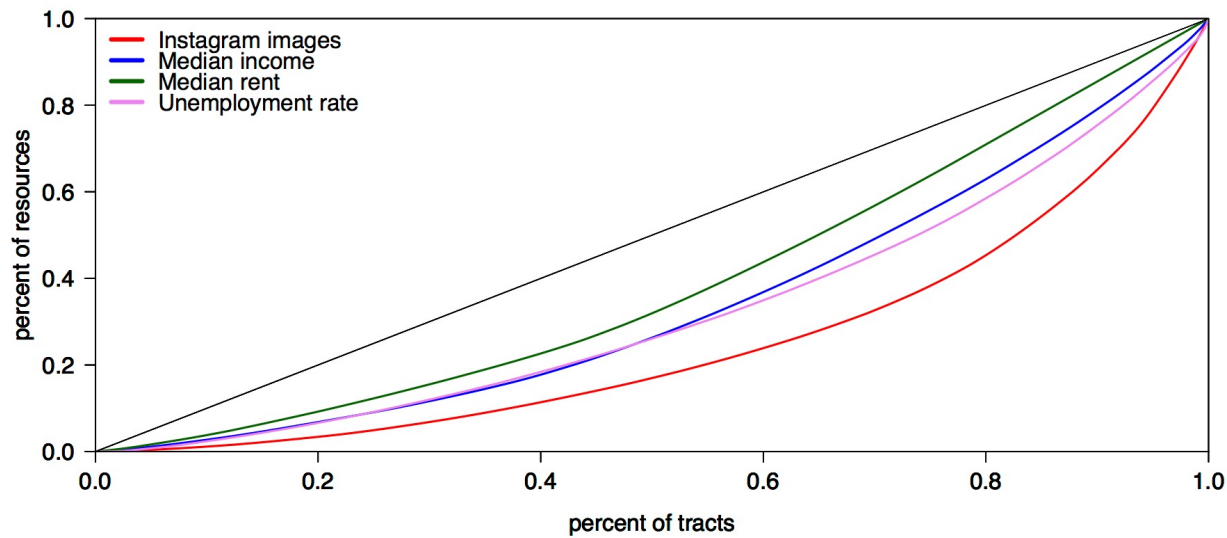

**Fig. 11.**

*The inequality of Instagram images shared in Manhattan turns out to be bigger than inequalities in levels of income, rent, and unemployment.* This is a very interesting and original result. Note that we are only considering images shared by local residents, which is what makes the comparison between distributions of social media and distributions of socio-economic indicators meaningful. We could have expected to see this result for visitors, given the concentration of most tourist landmarks and shopping areas in particular parts of the city. Finding that the inequality in Instagram shares is also larger than socio-economic inequality for local residents was really unexpected.

It is too early to draw big conclusions from this finding since we only looked at a single urban area (i.e., Manhattan). Nevertheless, recall that Manhattan has the highest income inequality among all urban areas in the U.S. (U.S. Census Bureau, 2014). Does this mean that in many other cities social media inequality will be even higher than inequality for socio-economic indicators? Or does it mean that social media signal amplifies already present social and economic inequalities in our societies? What are the relations between social media "portraits" of the cities created by postings of its residents and visitors, spatial patterns of socio-economic inequality, and locations of places of residence, work, and tourist attractions? Looking at data from many cities should help us answer these interesting questions.

# 5. Analysis of Social Media Inequality Using Hashtags

We can study social media inequality by analyzing counts and distributions of posts and users who share them. The examples are aggregated number of posts in a particular area from all users, average numbers of posts per user, or numbers of posts per hour or per day or the week.

We can also analyze *characteristics of social media content*. For example, we can calculate how topics of posts are distributed geographically. The topics can be extracted from text posts, from hashtags, and separately from images. We can ask if some geographic areas have more difference topics than other areas, and use inequality measures to quantify these differences.

Depending on the type of social media posts - texts only, text with links, images or video with descriptions, geo-located images, etc. - various existing computational methods can be used to quantify characteristics of content and analyze their spatial and/or temporal distributions. In this part of our paper we will illustrate this idea by identifying and counting the numbers of all hashtags assigned by users to the images they shared in Manhattan, and also numbers of unique hashtags.

Hashtags play a key role in Instagram experience. If a user adds hashtags to an image, that image becomes visible whenever all other app users search for any of these hashtags. Therefore, using hashtags also helps users reach a wider audience beyond their current followers. This was the initial intention of hashtags features: to increase visibility and expand the network of users of social media platforms. However, hashtags also have another important role – allowing users to describe their images and highlighting particular content they see important in these images. Given that an image may contain lots of different information, hashtags are ways of foregrounding some of this information. In other words, hashtags are a mechanism that assigns meaning to images.

Hashtags are particularly important for representation of cities on social media. People today use social media to research numerous places in a city – businesses, hotels, shops, cafes, restaurants, clubs, art galleries, museums, etc. – by looking at posts and images tagged with hashtags for these places. According to one report, "photo-heavy social media platforms have become more popular than review sites for sharing experiences and destination details."[5] The businesses and leisure destinations prominently display their hashtags in their windows or at the entrance, thus inviting visitors to share photos with these hashtags. A destination which has more photos on Instagram tagged with its hashtag becomes more important than a destination with fewer photos.

Hashtags and photos of destinations shared by people create a selective "map" of a city where some locations stand out from the rest, while some don't appear at all. This Instagram map competes with other online "maps" which, in contrast to traditional maps, are lists of places and events organized by categories and time rather than spatial relations. The examples are reviews of places in yelp.com, listings of events on eventful.com and eventbrite.com, and city portals such as afisha.ru and cult.mos.ru.

---

[5] Shankman (2014).

To isolate hashtags for our analysis, we considered all character strings preceded by # symbol. This method automatically captures hashtags in all languages. We have then calculated total numbers of tags and numbers of unique tags for every tract. The last statistics can be potentially quite useful. For example, if a tract has more unique hashtags that are legitimate words included in dictionary of a given language, this may indicate that the users who shared these images have richer vocabulary and potentially higher education levels.

|  | Locals | Tourists |
|---|---|---|
| Number of images | 5,918,408 | 1,524,046 |
| Number of all tags | 14,119,037 | 2,767,822 |
| Proportion of images with hashtags | 0.537 | 0.455 |
| Proportion of images with more than 5 tags / more than 10 tags | 0.125 / 0.048 | 0.087/ 0.031 |
| Mean number of tags per image | 2.385 2.78 (super locals) | 1.816 |
| Mean number of tags per image (excluding images that have no tags) | 4.439 4.7 (super locals) | 3.98 |

**Table 2.** Statistics for hashtags assigned to images by locals and tourists.

Table 2 presents statistics for hashtags assigned by locals and tourists to images they shared. As we can see, tourists on the average tag fewer images than locals (45.5% vs 53.7%). And for the images they tag, they assign fewer tags than locals (3.98 vs. 4.43).

When we calculated Gini coefficients for all tracts, new patterns emerge. Table 3 compares Gini measures for total hashtags, unique hashtags, and numbers of images. (As before, our coefficient calculations use aggregated numbers for each tract normalized by tract area.) For both total tags per tract and unique tags for tract, inequality is much larger for tourists than for locals.

|  | Locals | Visitors | Ratio between tourist and locals measures |
|---|---|---|---|
| Gini coefficient for images | 0.494 | 0.669 | 1.352 |
| Gini coefficient for tags | 0.514 | 0.678 | 1.318 |
| Gini coefficient for unique tags | 0.467 | 0.604 | 1.293 |

**Table 3.** Gini coefficients for hashtags assigned to images by locals and tourists.

To make this more concrete, consider that 50% of all hashtags by locals come from only 17% of the Census tracts in Manhattan. The ratio between a tract with highest number of tags and a tract with lowest number is 1:240. The same ratio for numbers of images is 1:173; for median family income it is 1:21.

The examples in this section only introduce the possibilities of inequality analysis that use characteristics of social media content — in this case, hashtags. It is possible to extend this analysis in many ways - for example, by analyzing the popularity of particular or top hashtags throughout the days, or in different geographic areas.[6]

Computational methods allow us to detect languages used in tags and image descriptions, and content of images. For example, the analysis of distribution of images of architecture in a city would be very relevant.

Even more importantly, in addition to statistics about content shared by people in a city and characteristics of this content, we can also study patterns in reception - i.e. how this content is used by other users both locally and globally. APIs of major social media services allow downloading numbers of likes and favorites for every piece of content, any comments, and also basic information (such as user name) for people who assigned these likes and favorites. This allows us to study relative social media popularity of cities, parts of cities, neighborhoods and particu-lar places. We can see how some places may rapidly become visible in a global landscape, and what developments are responsible for this.

---

[6] For an example of the analysis of neighborhood profiles using Instagram tags, see Tannen and State (2015).

For users of social media today, the world is no longer neatly organized in a spatial hierarchy of countries, big cities, neighborhoods and individual plates in these cities. We don't expect people to search Instagram for "China" or "France." If Google Maps allows us to explore the territories at many zoom levels, and see any detail in larger context, Instagram only offers photos (and a small percentage of videos) of concrete places and concrete situations. We can explore places by looking through photos tagged with their names. But it is impossible to see spatial and temporal patterns created by accumulations of all the photos shared, for example, in a larger city given present Instagram's user interface.

However, by downloading, analyzing, and visualizing these photos, along with their tags, descriptions, time stamps and geo coordinates, the researchers can piece together the collective "image of a city" and see how it changes over time. The concept of social media inequality and analytical methods explored in this paper allow us to measure these images, and compare them at arbitrary scales.

# 6. Conclusion

In this paper, we introduce the novel concept of social media inequality, which we define as the measures of distribution of characteristics of social media content shared in a particular geographic area or between areas. An example of such characteristics is the number of photos shared by all users of a social network such as Instagram in a given city or city area, or the content of these photos.

Using standard inequality measures used in other disciplines, we measure social media inequality for 7.4 million Instagram images posted in Manhattan. Regardless of what measure we use, we find that the distribution of Instagram posts is more unequal for visitors than for locals. We also find that the distribution of Instagram posts at the Census tract level is more unequal than that of income, rent and unemployment.

Using social media posts also allows us to learn about the temporal dynamics of Instagram sharing in Manhattan. By studying Instagram posts during the daytime and nighttime we find that differences in social media activity among parts of a city are to a large extent driven by commuting patterns.

We believe the concept of social media inequality introduced here can be extended in several ways. Analyzing other cities across the world may reveal that different relations between places of residence and work, income distributions, and tourist areas lead to different spatial and temporal patterns of social media sharing. This could pave the way for a new way to analyze and compare cities. There are also several other ways the data could be sliced ( for example by particular holidays or major events) to try to understand how city-life changes at particular moments. Finally, there are many other characteristics from social media posts that can be analyzed, not only from other social media platforms, but also by analyzing the distribution of likes or comments etc. throughout the city.

# Acknowledgements

The dataset used in this paper was originally created for *On Broadway* project (http://on-broadway.nyc) commissioned by New York Public Library. The project used only a part of the dataset: 661,809 Instagram images shared along Broadway. However, it allowed us to develop the conceptual framework for this paper: analysis social media patterns across small city areas, and comparison with Census indicators for these areas.

The original idea to compare images shared along Broadway comes from Daniel Goddemeyer. Daniel and Moritz Stefaner designed a number of visual interfaces for interacting with these images and other data sources. The case study in our paper builds on this work by expanding analysis to all of Manhattan.

Jay Chow downloaded all Instagram data and images. Mehrdad Yazdani created locals and visitors datasets, analyzed hashtags, and separated Instagram data according to Census boundaries working together with Ran Goldblatt. Ran created maps for figure 3.

*On Broadway* project team:

Artists:
Daniel Goddemeyer, Moritz Stefaner, Dominikus Baur, Lev Manovich.

Contributors:
Software Studies Initiative (Mehrdad Yazdani, Jay Chow), Brynn Shepherd and Leah Meisterlin, PhD students at The Graduate Center, City University of New York (Agustin Indaco, Michelle Morales, Emanuel Moss, Alise Tifentale).

*On Broadway* project uses software tools developed by Software Studies Initiative with the support from The Andrew W. Mellon Foundation and National Endowment for Humanities.

The project was supported by New York Public Library, The Graduate Center, City University of New York and California Institute for Telecommunications and Information at University of California, San Diego.